# An *sp*-hybridized molecular carbon allotrope, cyclo[18]carbon

Katharina Kaiser,[1†] Lorel M. Scriven,[2†] Fabian Schulz,[1] Przemyslaw Gawel,[2]* Leo Gross,[1]* Harry L. Anderson[2]*

[1] IBM Research–Zurich, Säumerstrasse 4, 8803 Rüschlikon, Switzerland

[2] Department of Chemistry, Oxford University, Chemistry Research Laboratory, Oxford, OX1 3TA, United Kingdom

[†] These authors contributed equally.

*Corresponding authors. Email: przemyslaw.gawel@chem.ox.ac.uk; lgr@zurich.ibm.com; harry.anderson@chem.ox.ac.uk

**Abstract:**

Carbon allotropes built from rings of two-coordinate atoms, known as cyclo[*n*]carbons, have fascinated chemists for many years, but until now they could not be isolated or structurally characterized, due to their high reactivity. We generated cyclo[18]carbon ($C_{18}$) using atom manipulation on bilayer NaCl on Cu(111) at 5 Kelvin by eliminating carbon monoxide from a cyclocarbon oxide molecule $C_{24}O_6$. Characterization of cyclo[18]carbon by high-resolution atomic force microscopy revealed a polyynic structure with defined positions of alternating triple and single bonds. The high reactivity of cyclocarbon and cyclocarbon oxides allows covalent coupling between molecules to be induced by atom manipulation, opening an avenue for the synthesis of other carbon allotropes and carbon-rich materials from the coalescence of cyclocarbon molecules.

The discovery of fullerenes (1), carbon nanotubes (2), and graphene (3), all of which consist exclusively of 3-coordinate carbon atoms, has sparked a new field of synthetic carbon allotropes (4, 5). The only molecular allotropes of carbon that have been isolated are the fullerenes (1). Rings of 2-coordinate carbon atoms (cyclo[$n$]carbons, $C_n$) have been suggested as an alternative family of molecular carbon allotropes (4-6). There is evidence for the existence of cyclocarbons in the gas phase (4, 5, 7, 8), but these highly reactive species have not been structurally characterized or studied in condensed phases. Whether cyclocarbons are polyynic, with alternating single and triple bonds of different lengths ($D_{9h}$ symmetry), or cumulenic with consecutive double bonds ($D_{18h}$ symmetry, see Fig. 1) is fundamental and controversial (9-12).

A distinctive feature of *sp*-hybridized carbon allotropes is that they possess two perpendicular π-conjugated electron systems (Fig. 1). Hückel's rule predicts for planar, cyclic conjugated systems with (4$n$ + 2) π electrons an aromatic structure with no bond length alternation (BLA) (13). Hoffmann predicted, in 1966, that two orthogonal ring currents would be formed in $C_{18}$, causing double aromatic stabilization (6). Since then, the structures of cyclo[$n$]carbons have been a topic of theoretical debate, and the results depend on the level of theory (9, 10, 13). Most density functional theory (DFT) and Møller-Plesset perturbation theory calculations predict that the lowest energy geometry of $C_{18}$ is cumulenic $D_{18h}$ (9, 10), whereas Hartree-Fock (4) and high-level Monte Carlo and coupled cluster methods predict that the polyyne $D_{9h}$ form is the ground state (11, 12).

Most attempts at synthesizing cyclocarbons have used a masked alkyne equivalent incorporated into a cyclic precursor designed to generate cyclo[$n$]carbon when activated by heat or light. Methods of unmasking included a retro-Diels-Alder reaction (4), decomplexation (14), decarbonylation (15) and [2+2] cycloreversion (16). These attempts have given tantalizing glimpses of cyclo[$n$]carbon in the gas phase. Cyclo[$n$]carbons may coalescence to form fullerenes (8, 15), and gas-phase electronic spectra of $C_{18}$, formed by laser ablation of graphite, have been measured, but these studies did not reveal whether the structures are cumulenic or polyynic (17).

An alternative approach for studying highly reactive molecules is to isolate them on an inert surface at low temperature. Developments in the field of scanning tunneling microscopy (STM) and atomic force microscopy (AFM), in particular tip functionalization, have enabled imaging of molecules with unprecedented resolution (18, 19), and atom manipulation can trigger chemical reactions on surfaces (19-22). We report the synthesis and structural characterization of cyclo[18]carbon. A low-temperature STM-AFM was used to sequentially remove masking CO groups from the precursor $C_{24}O_6$ by atom manipulation (Fig. 2). We resolved the structure of cyclo[18]carbon and observed BLA in its ground state. We also demonstrated covalent fusion by atom manipulation of cyclocarbon oxides, exploiting their high reactivity.

Cyclocarbon oxides, developed by Diederich and co-workers, were selected as suitable candidates for on-surface cyclocarbon generation (14). We synthesized the cyclocarbon oxide $C_{24}O_6$ using procedures similar to those previously reported (for synthetic details and NMR spectra see supplementary material and figs. S1-S7) and sublimed it from a Si wafer onto a cold ($T \approx 10$ K) Cu(111) surface partially covered with (100)-oriented bilayer NaCl islands. All molecules were studied on bilayer NaCl, providing an inert surface on which radicals and polyynes (22) are stable and can be imaged (19). The experiments were carried out in a combined STM/AFM system equipped with a qPlus force sensor (23) operating at $T = 5$ K in frequency-modulation mode (24). We used CO tip functionalization to improve the resolution (18). AFM images were acquired at constant height, with the offset $\Delta z$ applied to the tip-sample distance with respect to the STM setpoint above the bare NaCl surface. We simulated AFM images using the probe particle model (25) based on the structures relaxed in the gas phase, calculated by DFT (see also supplementary text).

After deposition, we found molecules of the precursor $C_{24}O_6$ on the NaCl surface, as well as some fragmented or (partially) decarbonylated molecules and single CO molecules (see supplementary material fig. S8 for an overview image). This result indicated that partial decarbonylation and dissociation took place during sublimation. Figure 3 shows AFM data and corresponding simulations for $C_{24}O_6$ and products created by atom manipulation (STM images are shown in fig. S9). $C_{24}O_6$ molecules appeared as triangular objects with dark features at the corners and two bright protrusions at each side (Fig. 3B). The dark contrast, characteristic of ketone groups (26) was explained by a reduced adsorption height and a relatively small electron density in the region imaged above the O atoms, which have high electron affinity. Both effects led to comparably small Pauli repulsive forces above the O atoms (26). Figure 3B was recorded at moderate tip height ("AFM far"), at which differences in bond order were visible in the $\Delta f$ signal, with high brightness, i.e., high $\Delta f$ indicating high bond order (27). The two bright features at each side of the molecule were assigned to triple bonds (22, 28). At decreased tip-sample distance ("AFM close", Fig. 3C), repulsive forces made greater contributions and tip relaxations, i.e., tilting of the CO at the tip apex, affected the AFM images substantially. These effects led to apparent sharpening of the bonds, and pronounced differences in apparent bond lengths (25, 27, 29). The different contrasts of the three different sides of $C_{24}O_6$ (Fig. 3B) suggest an adsorption geometry not parallel to the surface (26).

To decarbonylate $C_{24}O_6$, the tip was positioned in the vicinity (a few nanometers) of the molecule, retracted by about 3 Å from the STM setpoint (typically $V = 0.2$ V and $I = 0.5$ pA) and the sample bias voltage $V$ was increased to about $V = +3$ V, for a few seconds. This procedure often led to the removal of two, four, or six CO moieties. Because of the nonlocality and the observed bias thresholds (see supplementary text), we tentatively propose that the reaction was mediated by inelastic electron tunneling through hot interface state electrons (19). The most abundant products were $C_{22}O_4$ (Fig. 3F) and $C_{20}O_2$ (Fig. 3K). The removal of a masking group (2 CO) resulted in the formation of a longer bent polyynic segment.

Assigning the bright features in the "AFM far" images to the location of triple bonds, we observed curved polyyne segments with the expected number of triple bonds: 5 in $C_{22}O_4$ (Fig. 3G) and 8 in $C_{20}O_2$ (Fig. 3L). At small tip height (Fig. 3, H and M), we observed sharp bond-like features with corners at the assigned positions of triple bonds and straight lines in between. This contrast was explained by CO tip relaxation, in that maxima in the potential energy landscape, from which the tip apex was repelled, were located above the triple bonds because of their high electron density. In between these maxima, ridges in the potential landscape led to straight bond-like features (25, 30, 31). The assignment of the intermediates was further supported by AFM simulations (Fig. 3, 4$^{th}$ and 5$^{th}$ column) and STM images within the fundamental gap and at the ion resonances (see supplementary material figs. S10-S13).

We removed all six CO moieties from $C_{24}O_6$, with 13% yield (calculated by evaluation of 90 atomic manipulation events; for details on the statistics and procedure for on-surface decarbonylation, see supplementary text, table S1), typically resulting in circular molecules (Fig. 3, Q and R). At moderate tip heights (Fig. 3Q) the resulting molecule exhibited a cyclic arrangement of nine bright lobes. One side of the molecule appeared brighter, indicating that its adsorption geometry was not parallel to the surface (see supplementary fig. S14). For smaller tip heights (Fig. 3R), the molecule appeared as a nonagon with corners at the positions of the bright lobes that were observed at larger tip-sample distance (Fig. 3Q). The contrast can be explained in analogy to the precursors: The bright lobes in Fig. 3Q and the corresponding corners in Fig. 3R were observed above triple bonds. The molecule was thus identified as cyclo[18]carbon. The AFM contrast provided evidence for a polyynic structure of cyclo[18]carbon on NaCl with the defined positions of triple bonds supported by AFM simulations (Fig. 3, S and T and fig. S15). In the case of a $D_{18h}$ cumulenic structure, no BLA and an 18-fold symmetry of the molecule would be expected, in contrast to the experimental result.

In the ninefold symmetric form described above, the cyclo[18]carbon molecule was uncharged (neutral). We found that cyclo[18]carbon exhibited charge bistability (32) on this surface and it changed to a less symmetric and less planar geometry in the negatively charged state. The molecule could be reversibly switched between its two charge states and associated geometries (figs. S16-S18). During charge-state switching, tip-induced decarbonylation, and STM imaging, the molecules often jumped to different locations on the surface, indicating a very small diffusion barrier on NaCl. Cyclo[18]carbon moved more frequently than the other molecules observed, and it was often found adjacent to step edges or individual CO molecules adsorbed on the surface, pinning the molecule (Fig. 3, Q and R).

The high reactivity of cyclo[18]carbon and its oxides makes them promising candidates for on-surface covalent molecular fusion by atom manipulation, of which very few previous examples have been reported (20). Applying an elevated bias voltage near two proximate molecules led to their fusion. For example, the two neighboring cyclocarbon oxide intermediates, $C_{20}O_2$ and $C_{22}O_4$ (Fig. 4A), were fused in this way. After constant-current imaging at a set point of $V = 3$ V and $I = 0.5$ pA, the molecules reacted through a [4+2] cycloaddition, as revealed by the AFM resolved structure of the product shown in Fig. 4C (see fig. S19 for a possible mechanism and figs. S20-S21 for more examples of on-surface fused molecules). These results demonstrated that the strained polyynic moieties of cyclo[18]carbon and its oxide intermediates allowed covalent coupling by atom manipulation. Our results provide direct experimental insights into the structure of a cyclocarbon and open the way to create other elusive carbon-rich molecules by atom manipulation.

**Acknowledgments:** We thank Nikolaj Moll and Rolf Allenspach for discussions. **Funding:** We thank the ERC (grant 320969 and 682144) and the Leverhulme Trust (Project Grant RPG-2017-032) for support. P.G. acknowledges the receipt of a Postdoc.Mobility fellowship from the Swiss National Science Foundation (P300P2-177829). **Author contributions:** L.M.S., P.G. and H.L.A. synthesized the precursors; K.K., F.S., L.M.S. and L.G. performed the AFM experiments; F.S., P.G. and K.K. carried out the DFT calculations; F.S. performed the AFM simulations; all authors contributed to conceiving the research and writing the manuscript. **Competing interests:** The authors declare no competing interests. **Data and materials availability:** All data needed to evaluate the conclusions in the paper are available in the main text or the supplementary materials.

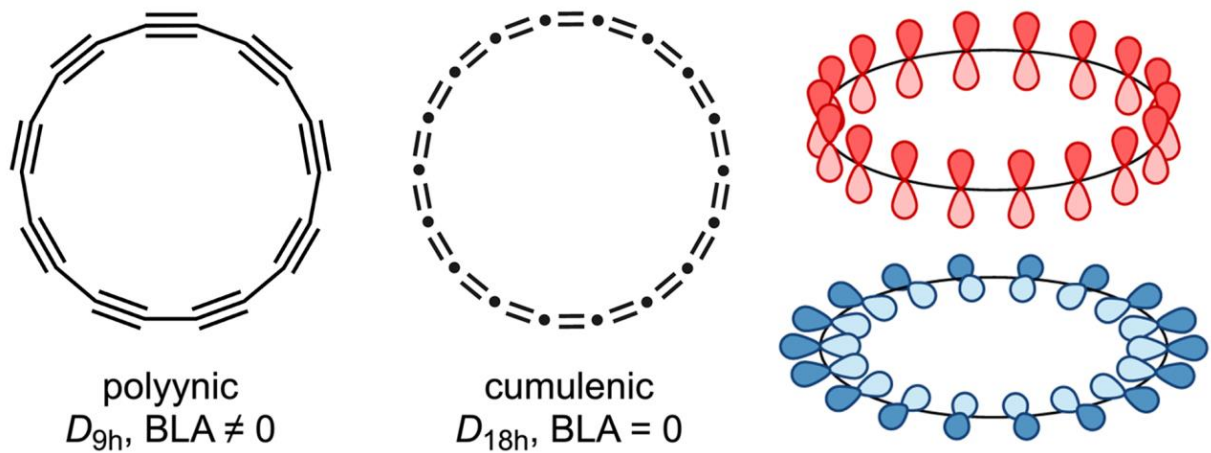

**Figure 1. Two possible structures of cyclo[18]carbon**: the polyynic form with $D_{9h}$ symmetry and the cumulenic form with $D_{18h}$ symmetry, (BLA = bond length alternation) and visualization of their perpendicular π-systems.

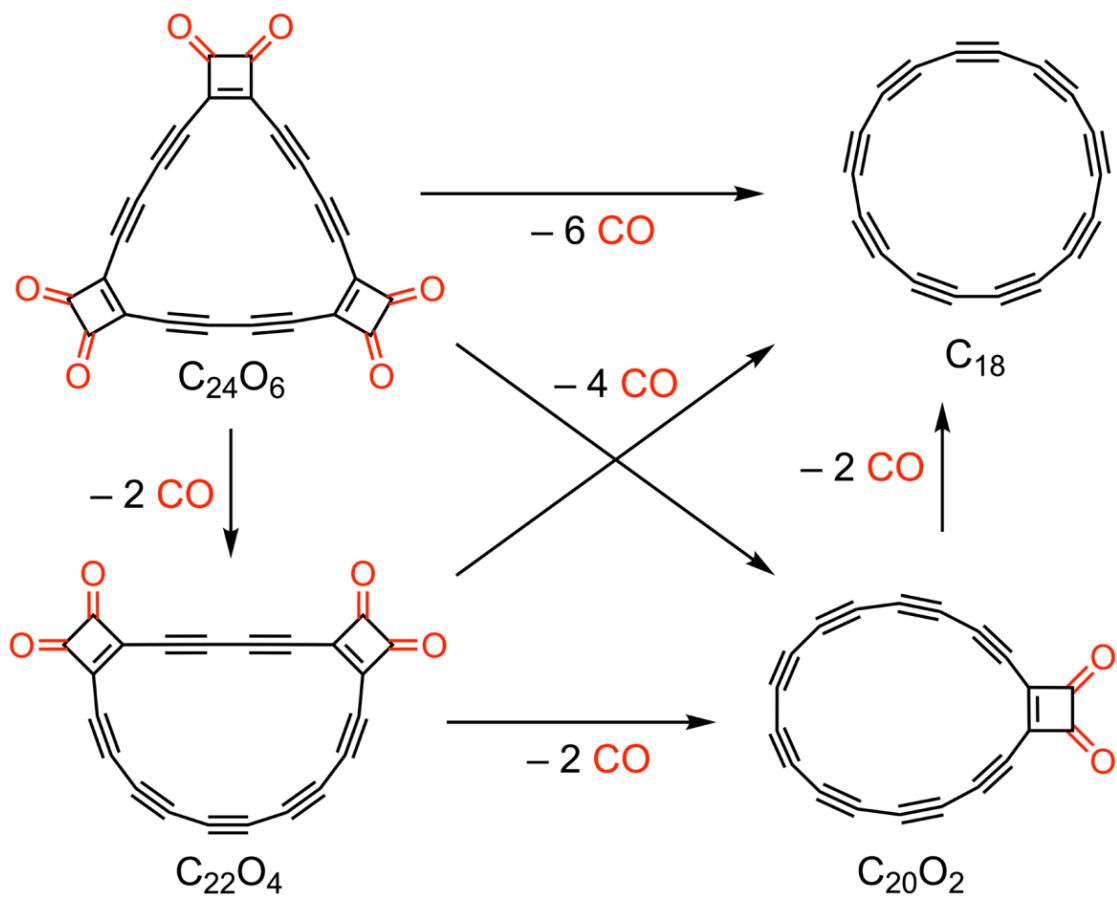

**Figure 2: Reaction scheme for the on-surface formation of C$_{18}$.** Decarbonylation was achieved *via* voltage pulses resulting in the loss of two, four or six CO moieties.

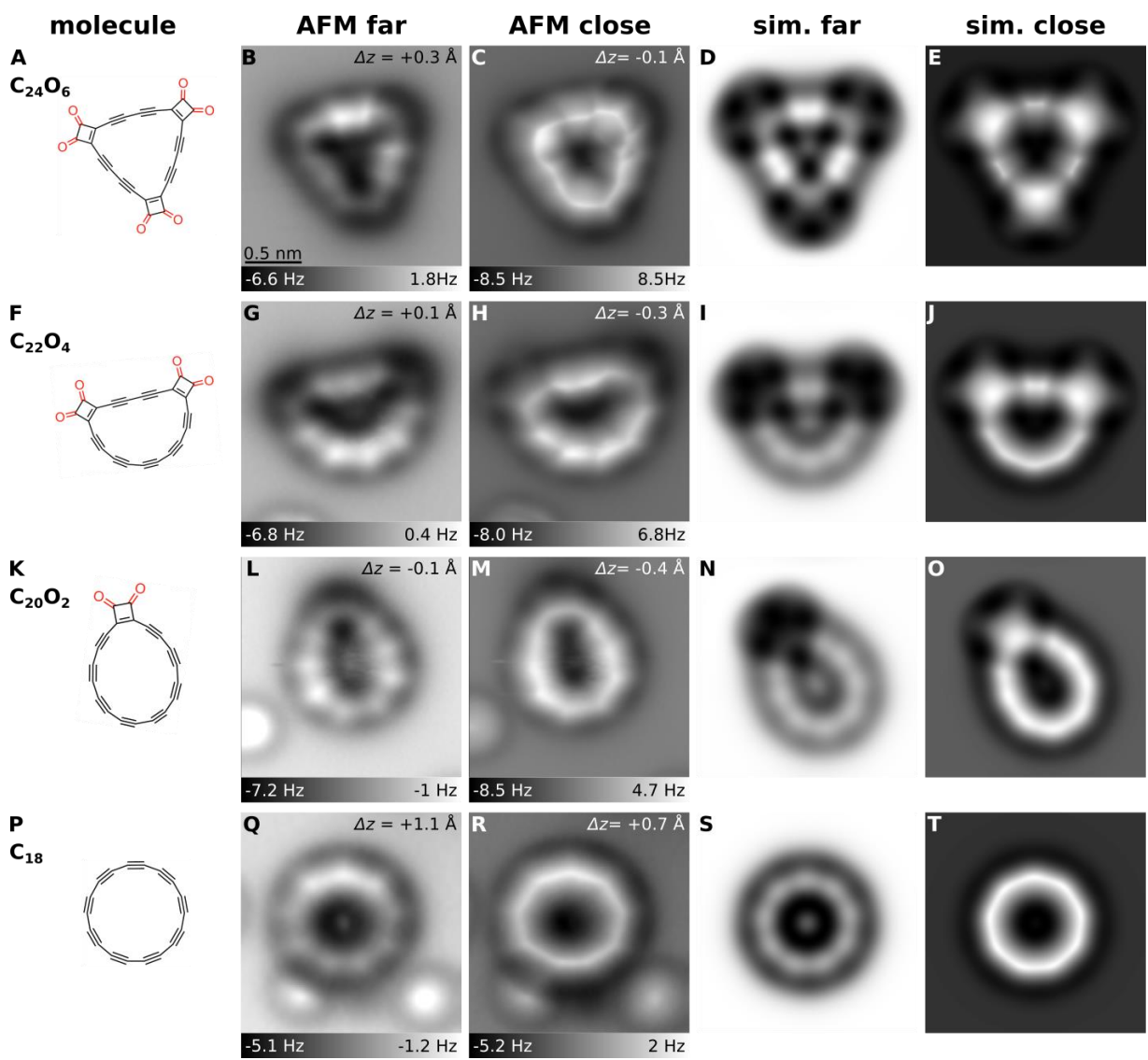

**Figure 3: Precursor and products generated by tip-induced decarbonylation.** Structures (1st column) and AFM images (2nd and 3rd column) recorded with a CO-functionalized tip at different tip offsets Δ$z$, with respect to an STM set point of $I$ = 0.5 pA, $V$ = 0.2 V above the NaCl surface, showing (**A**–**E**) precursor, (**F**–**J** and **K**–**O**) most frequently observed intermediates and (**P**–**T**) cyclo[18]carbon. The bright features at the lower part in L, M, Q and R correspond to individual CO molecules. Simulated AFM images (4th and 5th column) based on gas-phase DFT calculated geometries. The difference in probe height in "sim. far" to "sim. close" corresponds to the respective differences in "AFM far" and "AFM close". The same scale bar as in (B) applies to all experimental and simulated AFM images.

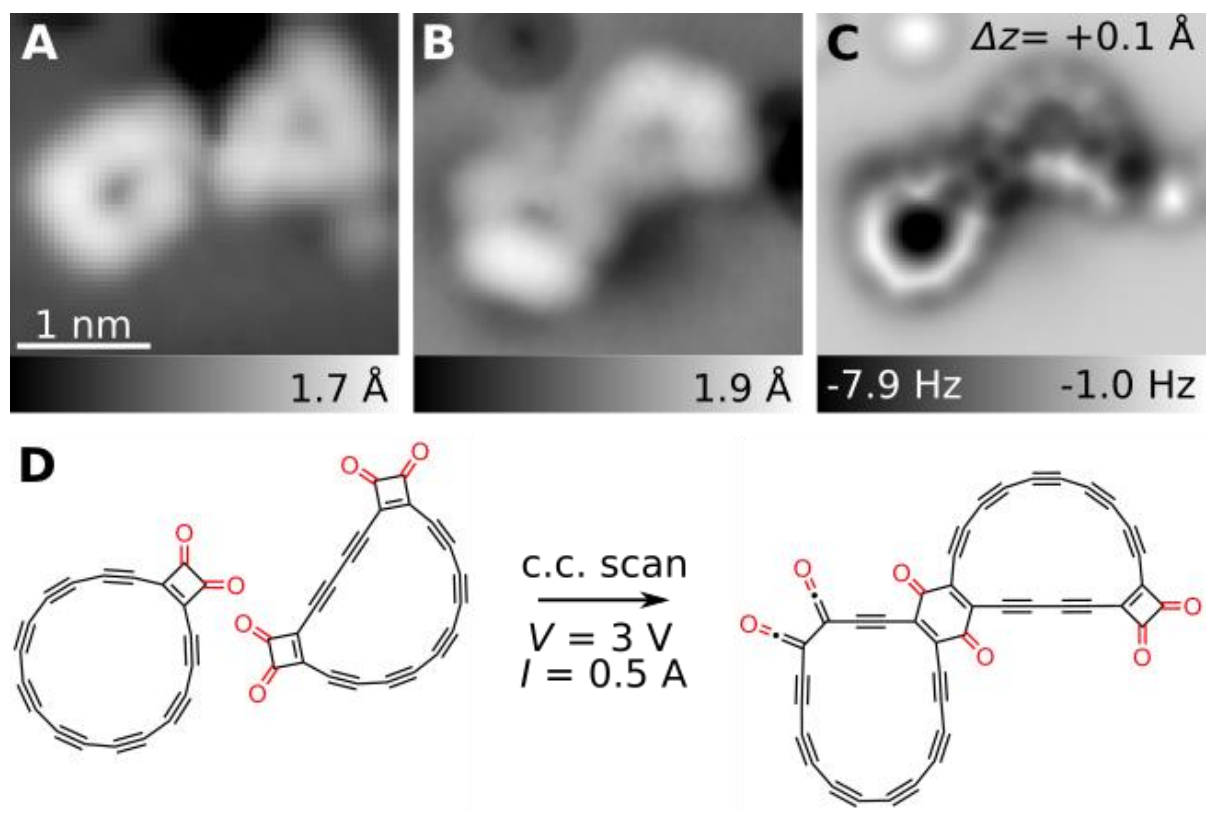

**Figure 4: Molecular fusion by atom manipulation**. (**A**) STM image of two neighboring intermediates, identified as $C_{22}O_4$ and $C_{20}O_2$. Imaging the area at constant current (c.c.) with a set point of $V = 3$ V, $I = 0.5$ pA resulted in the reaction between the molecules. (**B**, **C**) STM and AFM image of the resulting fused molecule, respectively. (**D**) Proposed reaction scheme. See supplementary material for further examples of molecular fusion reactions.

# Supplementary Materials for

## An *sp*-hybridized molecular carbon allotrope, cyclo[18]carbon

Katharina Kaiser,[1†] Lorel M. Scriven,[2†] Fabian Schulz,[1] Przemyslaw Gawel,[2*] Leo Gross,[1*] and Harry L. Anderson[2*]

[1] IBM Research–Zurich, Säumerstrasse 4, 8803 Rüschlikon, Switzerland

[2] Department of Chemistry, Oxford University, Chemistry Research Laboratory, Oxford, OX1 3TA, United Kingdom

[†] These authors contributed equally.

[*]Corresponding authors. E-Mail: przemyslaw.gawel@chem.ox.ac.uk; lgr@zurich.ibm.com; harry.anderson@chem.ox.ac.uk

**This PDF file includes:**

Materials and Methods
Supplementary Text SM1
Table S1
Figs. S1 to S21
Reference 33 to 51

## Materials and Methods

### Sample and Tip Preparation

As a substrate, a Cu(111) single crystal was partially covered with NaCl. The Cu single crystal was cleaned *in situ* by repeated Ne-ion sputtering and annealing ($T \approx 720$ K) cycles. NaCl was sublimed onto the clean Cu surface that was kept at $T \approx 270$ K, resulting in (100)-oriented, mostly double-layered NaCl islands (33). The microscope tip (PtIr-wire with a diameter of 25 μm) was sharpened using a focused ion beam, followed by *in situ* indentations into the Cu surface to prepare an atomically sharp tip. For tip functionalization, we used single CO molecules that were picked up from the surface (18, 34). To this end, gaseous CO was allowed into the UHV chamber (partial pressure of $p = 2 \times 10^{-8}$ mbar) for about 20 seconds to adsorb on the cold ($T \approx 10$ K) surface. The cyclocarbon oxide precursors were deposited on the cold ($T \approx 10$ K) sample *via* sublimation from a Si-wafer that was flash-annealed to approximately 900 K within a few seconds (35). An exemplary overview image is shown in fig. S8.

### STM/AFM Measurements

For the characterization with STM and AFM, we used a home-built combined STM/AFM system, operating at ultra-high vacuum conditions ($p \approx 1 \cdot 10^{-10}$ mbar) and low temperatures ($T = 5$ K). The microscope employs a qPlus force sensor (23) with a resonance frequency $f_0 \approx 30$ kHz, quality factor $Q \approx 100.000$ at $T = 5$ K and a spring constant $k \approx 1.800$ Nm$^{-1}$ and is operated in frequency-modulation mode (24). The bias voltage $V$ was applied to the sample. The lower end of the tip-height color scale bar (corresponding to black) in constant-current STM images is defined as 0 Å. AFM images were acquired in constant-height mode at $V = 0$ V and an oscillation amplitude of $A = 0.5$ Å. The tip-height offsets $\Delta z$ for constant-height AFM images are defined as the offset in tip-sample distance relative to the STM set point at the top center of the acquired image, with positive (negative) values indicating that the tip-sample distance increased (decreased) with respect to the STM set point.

### Synthetic General Methods

Reagents (Alfa Aesar, Aldrich, Acros, Fluorochem, Fisher Scientific) were used without further purification. Dry solvents (THF, Et$_2$O) for reactions were purified by a MBraun MB-SPS-5-Bench Top under nitrogen (H$_2$O content < 20 ppm). All other solvents used were HPLC grade. Reactions, unless otherwise stated, were carried out in oven-dried glassware under a N$_2$ atmosphere. Flash column chromatography was carried out on a Biotage Isolera One with a 200–400 nm UV detector. Analytical thin layer chromatography (TLC) were performed on aluminum sheets coated with silica gel 60 F254 (Merck). UV light (254 nm) was used for visualization. Evaporation *in vacuo* was performed at 15–40 °C and 5–1010 mbar. Reported yields refer to pure compounds dried under high vacuum (< 0.1 mbar). $^1$H and $^{13}$C nuclear magnetic resonance (NMR) were recorded on Bruker AVIII HD 400 spectrometer at 400 MHz ($^1$H) and 101 MHz ($^{13}$C) at 21 °C. Chemical shifts, δ, reported in ppm downfield from tetramethylsilane using residual deuterated solvent signals as internal reference (CDCl$_3$: δ$_H$ = 7.26 ppm, δ$_C$ = 77.16 ppm; C$_6$D$_5$NO$_2$: δ$_C$ = 148.50 ppm). High-resolution mass spectrometry (HR-MS) measurements were performed by the mass spectrometry service at the University of Oxford on a Waters GTC classic.

## Synthetic Protocols

Cyclo[18]carbon oxide **S6** was prepared using modified procedures developed by Rubin, Diederich and coworkers as shown in Supplementary Scheme 1 (15, 36).

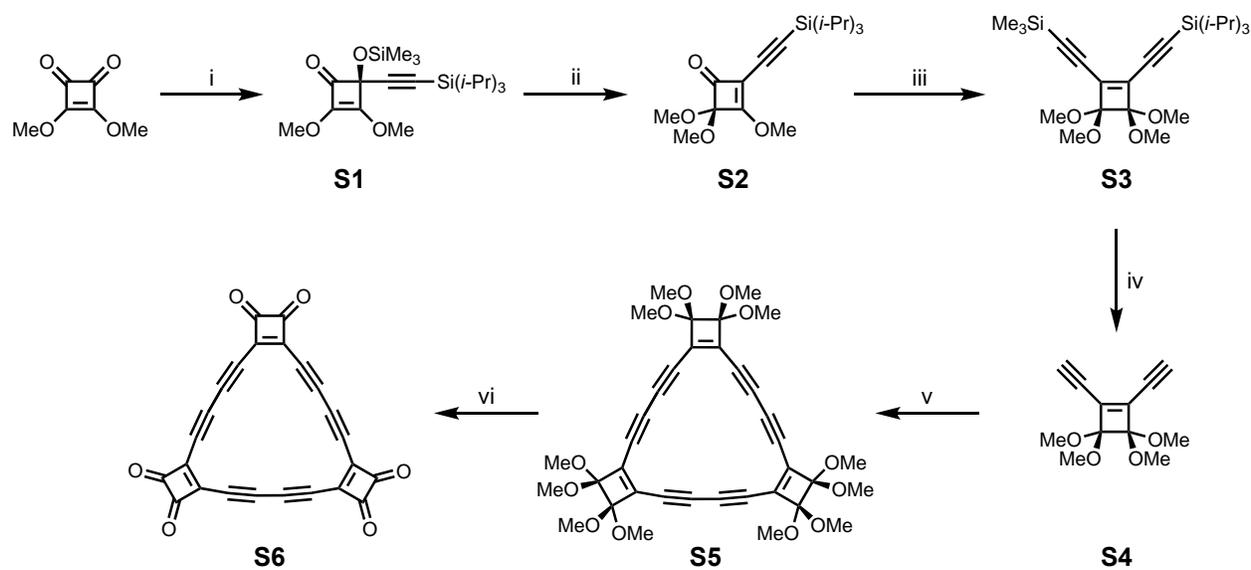

**Supplementary Scheme 1:** Synthesis of cyclo[18]carbon oxide **S6**. i) Li-C≡C-Si(*i*-Pr)$_3$, THF, 1 h, –78 °C; then TMS-Cl, 15 min, –78 °C to 21 °C, 96%; ii) MeOTMS then TMSOTf, THF, 30 min, 21 °C, 89%; iii) Li-C≡C-SiMe$_3$, Et$_2$O, 30 min, –45 °C; then HCl (3 M), –45 °C to 21 °C, 1 h; then MeOTMS then TMSOTf, 60 h, 21 °C, 75%; iv) TBAF, THF, 30 min, –78 °C, 93%; v) CuCl, TMEDA, acetone, O$_2$, 2 h, 21 °C, 7%; vi) conc. H$_2$SO$_4$, (CH$_2$Cl)$_2$, 10 min, 21 °C, 96%. TMS = trimethylsilyl; TIPS = triisopropylsilyl; Tf = triflate; TBAF = tetra-*n*-butylammonium fluoride; TMEDA = *N,N,N',N'*-tetramethylethylenediamine.

**Synthesis of compound S1:** *n*-BuLi (4.6 mL, 1.6 M in hexanes, 7.4 mmol) was added to a solution of TIPS-acetylene (1.7 mL, 7.4 mmol) in THF (8 mL) at 0 °C. The clear solution was stirred for 30 minutes and then added over 15 minutes to a dimethyl squarate (1.0 g, 7.0 mmol) solution in THF (30 mL) at –78 °C. After stirring for 1 h at –78 °C, TMS-Cl (1.2 mL, 9.2 mmol) was added and the solution was allowed to warm to 21 °C. The reaction was quenched with a sat. aqueous solution of NaHCO$_3$ (50 mL) and extracted with CH$_2$Cl$_2$ to yield **S1** as a yellow oil (2.7 g, 6.8 mmol, 96%). **$^1$H NMR** (400 MHz, CDCl$_3$): 4.17 (s, 3H; **a/b**), 3.95 (s, 3H; **a/b**), 1.07 (m, 21H; **d**), 0.25 ppm (s, 9H; **c**). **$^{13}$C NMR** (101 MHz, CDCl$_3$): 180.3, 165.7, 135.3, 102.6, 91.5, 79.8, 59.8, 58.6, 18.7, 11.3, 1.4 ppm. $^1$H and $^{13}$C NMR match that of the literature (36).

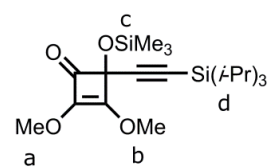

**Synthesis of compound S2:** MeOTMS (0.98 mL, 7.1 mmol) was added to a solution of **S1** (2.7 g, 6.8 mmol) in THF (3 mL). TMSOTf (73 µL, 0.41 mmol) was added dropwise to avoid overheating (an exothermic reaction occurs). The mixture was stirred for 30 minutes at 21 °C and then quenched 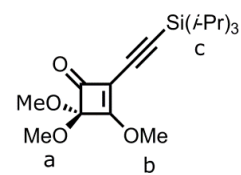 with a sat. aqueous solution of NaHCO$_3$ (5 mL). The aqueous phase was extracted with CH$_2$Cl$_2$ and the organic layer was dried over MgSO$_4$ to yield **S2** as a yellow oil (2.0 g, 6.0 mmol, 89%). **$^1$H NMR** (400 MHz, CDCl$_3$): 4.37 (s, 3H; **b**), 3.52 (s, 6H; **a**), 1.08 ppm (s, 21H; **c**). **$^{13}$C NMR** (101 MHz, CDCl$_3$): 186.4, 184.2, 112.6, 99.8, 95.2, 92.1, 61.3, 53.5, 18.7, 11.4 ppm. $^1$H NMR and $^{13}$C NMR match that of the literature (36).

**Synthesis of compound S3:** *n*-BuLi (4.0 mL, 1.6 M in hexanes, 6.4 mmol) was added to a TMS-acetylene (0.88 mL, 6.4 mmol) solution in Et$_2$O (15 mL) at 0 °C and stirred for 30 minutes. The resulting clear solution was cooled to –78 °C and a solution of **S2** (2.0 g, 6.0 mmol) in 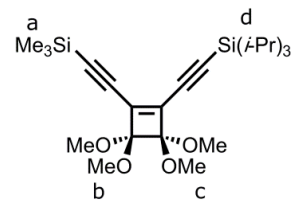 Et$_2$O (25 mL) was added dropwise over 30 minutes and the resulting solution was stirred for 1 h at –45 °C to give a dark red solution. Aqueous HCl (3 M, 20 mL) was added and the solution was stirred vigorously at 25 °C for 1 h. The reaction was quenched with a sat. aqueous solution of NaHCO$_3$ (50 mL) and the resulting organic layer dried over MgSO$_4$ to yield the enediyne as a yellow oil (2.39 g) which was used without further purification. MeOTMS (2.9 mL, 21.3 mmol) was added to enediyne (2.39 g) followed by addition of TMSOTf (0.13 mL, 0.74 mmol). The resulting brown solution was stirred for 60 h at 21 °C. The reaction was quenched with a sat. aqueous solution of NaHCO$_3$ (5 mL). The aqueous phase was extracted with CH$_2$Cl$_2$ and the organic layer was dried over MgSO$_4$ and purified by column chromatography, SiO$_2$ (petroleum ether/ethyl acetate 8:2) to yield **S3** as a yellow oil (2.0 g, 4.5 mmol, 75% over two steps). **$^1$H NMR** (400 MHz, CDCl$_3$): 3.51 (s, 6H; **c/b**), 3.50 (s, 6H; **b/c**), 1.08 (s, 21H; **d**), 0.18 ppm (s, 9H; **a**). **$^{13}$C NMR** (101 MHz, CDCl$_3$): 136.3, 136.0, 110.1, 108.3, 108.2, 107.5, 98.1, 96.0, 52.2, 52.1, 18.7, 11.3, –0.3 ppm. **HR-ESI-MS m/z**: 419.24251 (mass corresponds to C$_{24}$H$_{42}$O$_4$Si$_2^+$: 419.24322) **IR (ATR): ṽ** = 678, 761, 846, 883, 896, 919, 985, 1039, 1081, 1179, 1206, 1258, 1384, 1463, 1771, 2138, 2836, 2866, 2943 cm$^{-1}$.

**Synthesis of compound S4:** Tetrabutylammonium fluoride (2.2 mL, 1.0 M in THF, 2.2 mmol) was added to a solution of **S3** (1.0 g, 2.2 mmol) in THF (40 mL) and water (0.5 mL) at –78 °C. The resulting dark red solution was then stirred for 30 minutes. The reaction was quenched with water and the aqueous phase 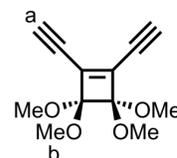 extracted with Et$_2$O (50 mL). The organic layer was washed with brine and dried over MgSO$_4$ to yield the crude product as a red oil. Purification by column chromatography, SiO$_2$ (petroleum ether/ethyl acetate 7:3) gave **S4** (456 mg, 2.05 mmol, 93%) as an unstable red crystalline solid that was used immediately in the next step. **$^1$H NMR** (400 MHz, CDCl$_3$): 3.73 (s, 2H; **a**), 3.50 (s, 12H; **b**) ppm. **$^{13}$C NMR** (101 MHz, CDCl$_3$): 136.3, 108.2, 91.2, 74.9, 52.1 ppm. **HR-ESI-MS:** 245.07852 (mass corresponds to C$_{12}$H$_{14}$O$_4$Na$^+$: 245.07843). **IR (ATR): ṽ** = 650, 812, 870, 983, 1030, 1078, 1179, 1199, 1257, 1440, 2103, 2838, 2944, 3243 cm$^{-1}$.

**Synthesis of compound S5:** Freshly prepared CuCl (37) (5.0 g, 51 mmol) was stirred in acetone (50 mL) under N$_2$. TMEDA (2.7 mL, 18 mmol) was added to this suspension and the resulting opaque, pale blue solution was stirred for 30 minutes. The solid was allowed to settle and the supernatant solution (15.0 mL, 5.29 mmol) 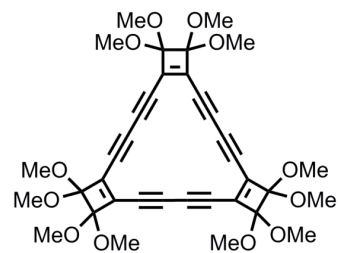 withdrawn and immediately added to a vigorously stirred solution of **S4** (1.15 g, 5.17 mmol) in acetone (300 mL) at 21 °C with a CaCl$_2$ drying tube and under an O$_2$ atmosphere for 2 h. Upon completion the reaction was diluted with CHCl$_3$ (300 mL) and the organic phase washed with water (2 × 600 mL). The organic extracts were combined and dried over MgSO$_4$ and the solvent removed *in vacuo* at 30 °C. The crude was purified by column chromatography, SiO$_2$ (petroleum ether/ethyl acetate 8:2) to yield **S5** as a yellow solid (76 mg, 0.12 mmol, 7%). **$^1$H NMR** (400 MHz, CDCl$_3$): 3.70 (s, 36H) ppm. **$^{13}$C NMR** (101 MHz, CDCl$_3$): 139.7, 109.9, 90.3, 82.1, 52.6 ppm. **MALDI-TOF m/z:** 629.308, (mass corresponds to C$_{35}$H$_{33}$O$_{11}^+$: 629.202). **IR (ATR): ṽ** =

625, 881, 982, 1034, 1083, 1180, 1209, 1252, 1457, 1560, 1742, 2836, 2940 cm$^{-1}$. **M.p.:** decomposes at ≈ 200 °C. **UV/vis (CHCl$_3$):** $\lambda_{max}$ (ε) = 403 (16100), 393 (12900), 345 (78100), 328 nm (42200 M$^{-1}$ cm$^{-1}$).

**Synthesis of compound S6:** All reaction vessels were kept under nitrogen and wrapped with aluminum foil to exclude light. Concentrated H$_2$SO$_4$ (0.2 mL) was dropped onto **S5** (10 mg, 0.015 mmol) to form a dark orange emulsion which was stirred for 5 minutes at 21 °C. Then, (CH$_2$Cl)$_2$ (10 mL) was added and the resulting solution stirred vigorously for 10 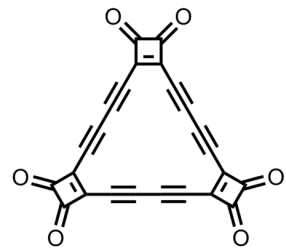 minutes. The clear orange organic phase was transferred *via* syringe to a flask containing CaCO$_3$ (0.4 g). The extraction with (CH$_2$Cl)$_2$ was repeated until the organic extracts were colorless. The combined organic phases were filtered with 0.45 µm PTFE syringe filters to yield a clear orange solution. The solvent was removed *in vacuo* and crude material was dried under high vacuum in the dark to yield **S6** as an unstable, light sensitive yellow solid which was stored in the dark at –20 °C (5.6 mg, 0.014 mmol, 96%). **$^{13}$C NMR** (101 MHz, C$_6$D$_5$NO$_2$): 84.5, 107.7, 180.7, 192.3 ppm. **IR (ATR):** $\tilde{\nu}$ = 804, 1028, 1103, 1403, 1541, 1788 (strong), 2117 cm$^{-1}$. **M.p.:** decomposes at ≈ 200 °C (Note this compound has been previously reported to explode violently above 85 °C). **UV/vis ((CH$_2$Cl)$_2$):** $\lambda_{max}$ (ε) = 438 (19300), 430 (19300), 381 (98600), 360 (50400), 355 (50400), 335 (26300), 316 (15200), 271 nm (31900 M$^{-1}$ cm$^{-1}$). $^{13}$C NMR, IR and UV-Vis spectra match that of the literature (15).

# Supplementary Text SM1

**Additional AFM data of C$_{18}$**

Figure S14 shows additional AFM images of the fully decarbonylated cyclo[18]carbon at different tip-height offsets $\Delta z$. For large tip-sample distances (fig. S14A), the molecules show bright features (increased $\Delta f$ values) that can be attributed to the triple bonds due to their large exposed electron density. For decreasing tip-sample distances, the brightness contrast between single and triple bonds gradually disappears and bright, bond-like features appear, with corners above the positions of triple bonds and straight lines connecting them (fig. S14D). As explained in the main text, this contrast results from the tilting of the CO molecule at the tip apex (25, 27, 30, 31, 38) and leads to the nonagon shape of the molecule. The AFM contrast indicates that cyclo[18]carbon molecules adsorbed in a non-planar geometry, i.e., not parallel with respect to the NaCl surface. The maxima related to the triple bonds appear with different brightness. From the different contrast of the triple bonds within AFM images and the contrast evolution as a function of tip height, we estimate maximal height differences of 0.1 Å to 0.3 Å within an individual molecule (26). Thus, we estimate tilts of the molecular plane with respect to the surface plane of about 1–2 degrees, varying for different molecules. The non-planar adsorption could be related to the mismatch in symmetry of the nine-fold symmetric cyclo[18]carbon with the fourfold symmetric NaCl surface. Such slight tilt of the molecular plane with respect to the surface can also be observed for C$_{24}$O$_6$ as well as for most of the intermediates (see Fig. 3 of main text).

The AFM images shown in fig. S14, C and D are taken at tip sample distances where a part of the molecule shows the contrast for small tip heights (cf. "AFM close" in Fig. 3 of main text), i.e., sharp lines with corners above the positions of triple bonds, while on the opposing side of the ring the contrast for large tip heights is visible ("AFM far" in Fig. 3 of main text), i.e., bright lobes above positions of triple bonds. These images visualize the non-planar adsorption geometry of the molecules and the evolution of the contrast as a function of tip-sample distance. Moreover, they corroborate that the corners observed at small tip height are at the locations of the bright lobes at large tip height, and thus correspond to the locations of the triple bonds.

Cyclo[18]carbon was very mobile on the surface and often jumped to a different location during STM and AFM imaging, indicating a low diffusion barrier on bilayer NaCl.

Cyclo[18]carbon was usually found adjacent to other adsorbates, mostly CO molecules. This is visible in fig. S14, A to C, where the bright contrast directly next to the cyclo[18]carbon stems from adjacently adsorbed CO molecules. We conclude that these CO molecules were not covalently bound to cyclo[18]carbon, due to the following observations: i) The adjacent CO molecules show the same contrast and adsorption site (on top of Na cations) as single isolated CO molecules, indicating they are also adsorbed in a vertical geometry on the NaCl surface. ii) If a cyclo[18]carbon jumped to a different adsorption site, the previously adjacent CO molecules did not move in conjunction with it. iii) The typical observed distance between cyclo[18]carbon and adjacent CO molecules, determined from the separation of $\Delta f$ maxima in AFM images (fig. S14, A to C), is about 4 Å, which is too large for a covalent bond, even when taking into account possible image distortions for tip relaxations. Presumably, the CO molecules interact with the cyclo[18]carbon *via* van der Waals forces and stabilize the cyclo[18]carbon molecule such that it can be imaged without moving. However, the presence of the closely adsorbed CO molecules can affect the tilting of the CO tip and thus lead to distortions during constant-height imaging at small tip-sample distances, affecting also the contrast observed above cyclo[18]carbon in the vicinity of CO molecules.

Figure S14E shows a constant-height AFM image in which the tip height was changed to determine the adsorption position of cyclo[18]carbon. In the top part of the image, a small tip-height offset of $\Delta z = -1.2$ Å (setpoint $V = 0.1$ V and $I = 0.5$ pA) was chosen to atomically resolve the NaCl lattice. The bright features in this part of the image correspond to Cl-atoms (39). In the part of the image below the green line the tip-height offset was increased to $\Delta z = +0.8$ Å. Extrapolating the Cl lattice sites, we find that this cyclo[18]carbon is adsorbed with the center on a Cl-Cl bridge site. At the red line the CO-functionalized tip lost the CO.

**STM data**

For the characterization of the electronic structure, STM images were recorded within the fundamental gap (denoted as in-gap), see fig. S9, and at the positive and negative ion resonances (PIR/NIR) for the precursor and the two most abundant intermediates, see fig. S10. For the $C_{24}O_6$ precursor, the in-gap STM image (fig. S9C) shows a bright feature in the center of the upper left side. This protrusion in the in-gap STM image was found in all imaged $C_{24}O_6$ molecules. It always corresponded to the bright side visible in the AFM images and therefore probably relates to the side of the molecule that has the largest adsorption height. For cyclo[18]carbon, the in-gap STM shows four-fold symmetry with four protrusions, leading to a square-like shape (fig. S9L). The diagonals of the square were always aligned with non-polar directions of the NaCl(100) surface. The four-fold contrast we assign to the influence of the underlying four-fold symmetric NaCl surface.

The orbital density images of precursor and intermediates (fig. S10) show qualitatively different contrast and different nodal planes for PIR and NIR, indicating that these molecules are in a closed shell configuration and thus charge neutral when imaging within the gap. We thus can assign the STM image recorded at the PIR to the density of the highest occupied molecular orbital (HOMO, recorded at negative bias voltage) and the NIR to the density of the lowest unoccupied molecular orbital (LUMO, at positive bias voltage). For cyclo[18]carbon, imaging of the ion resonances was not possible due to its charge bistability. Charge-state switching to the negative charge state was observed for positive bias, $V > 0.6$ V. At negative bias we did not observe a resonance for cyclo[18]carbon up to sample voltages of $V < –1.5$ V and the molecule was not stable (jumped on the surface) for increased negative voltages.

Figure S12 shows a cyclo[18]carbon adsorbed directly next to a $C_{24}O_6$ precursor molecule. Both molecules are not covalently linked, as is apparent from the AFM images at different tip-sample distances (fig. S12, A to C). At $V = –0.8$ V ($V = +1.3$ V), the onset of the PIR (NIR) of $C_{24}O_6$ is reached, leading to an increase in tunneling current at this voltage at the positions of large densities of the HOMO (LUMO) above the $C_{24}O_6$ molecule in constant-height STM images (fig. S12, E to F). No contrast (no tunneling current) is measured above cyclo[18]carbon in these images.

**The charge state of cyclo[18]carbon on double-layer NaCl on Cu(111)**

The charge state of a molecule can crucially influence its structure (40). The charge state of molecules can be inferred *via* observation of the standing wave patterns of interface-state electrons (41, 42). On NaCl/Cu(111), the Cu(111) surface state survives as an interface state. Scattering patterns of interface-state electrons can be made visible by imaging at low bias voltage (43). The scattering of interface-state electrons serves as a hallmark for charged particles because charged adsorbates act as scattering centers, but neutral adsorbates do not (32, 41). Figure S16A shows an STM image with a standing wave pattern of the interface state visible.

Because of scattering at step edges and island edges (43) it is not obvious if the cyclo[18]carbon molecule in the center of the image acts as a scattering center. The molecule was moved laterally, and the same area was imaged again, using the same set point (fig. S16B). To emphasize changes in the scattering pattern, caused by moving the molecule, both images were subtracted from each other (fig. S16C). For a charged molecule a pronounced standing wave pattern would be expected in the difference image. The resulting difference image does not exhibit such a standing wave pattern, which evidences that this cyclo[18]carbon is not acting as a scattering center for interface-state electrons and thus is charge neutral.

However, cyclo[18]carbon can be charged and exhibits charge bistability on this surface. This behavior is demonstrated in fig. S17. A change in the charge state of the molecule was deliberately triggered by STM imaging at $V = +0.6$ V and is evidenced by the change in standing wave pattern of the interface-state electrons. Figure S17A shows an STM image of a neutral cyclo[18]carbon, acquired at a set point of $I = 0.4$ pA, $V = 0.1$ V. By imaging the same molecule at a bias voltage of $V = +0.6$ V (fig. S17B), an abrupt change in shape occurred, during the scan line indicated by the arrow. Subsequent imaging of the same area, shown in fig. S17C, recorded with the same set point as in fig. S17A, shows a change in the standing wave pattern around the molecule compared to fig. S17A. This change is emphasized in the difference image fig. S17D, where the interference pattern around the molecule indicates a change in charge state between fig. S17A and fig. S17C.

Figure S18 shows a series of charging and discharging events of an individual cyclo[18]carbon molecule. The molecule was charged by imaging it at voltages $V > 0.6$ V. Presumably, cyclo[18]carbon is charged negatively by resonantly attaching a long-lived electron to its LUMO, in analogy to the reported charge-state switching of CuPc (32). In contrast to neutral cyclo[18]carbon, the charged molecule exhibits a distorted, less circular geometry (see fig. S18, D, E and H). Its structural characterization is even more challenging than for neutral cyclo[18]carbon, as it often changed its adsorption geometry and site, leading to the stripy images observed in fig. S18, D, E and H. At voltages $V < -0.2$ V, the molecule was switched back to the neutral charge state reestablishing the $D_{9h}$ geometry of neutral cyclo[18]carbon.

Often a change in the charge state of the molecule was accompanied by lateral displacement of the molecule and sometimes chemical reactions with surrounding molecules.

The charge bistability and the high reactivity render imaging of the orbital densities of cyclo[18]carbon with STM extremely challenging. As a result of the charging of cyclo[18]carbon at positive bias voltages, the LUMO related resonance cannot be spatially imaged. The threshold for charging the molecule at about $V = 0.6$ V gives an indication for the onset of the NIR, related to the LUMO, at this voltage. Once the molecule has been charged negatively, the corresponding state remains occupied and is therefore not available as a tunneling path. The next-higher unoccupied state of the singly negatively charged molecule lies higher in energy by the Coulomb energy and can therefore only be addressed by applying markedly higher voltages. Applying voltages larger than 1.5 V for imaging of cyclo[18]carbon has resulted in this molecule reacting with other molecules on the surface in all attempts. Imaging the HOMO density at negative bias voltages was also not possible since imaging at elevated negative bias voltages ($V < -1.5$ V) led to the molecule jumping away.

## On-surface synthesis of cyclo[18]carbon and reaction statistics

For the on-surface synthesis of cyclo[18]carbon, the tip was positioned in the vicinity of a precursor molecule and held at that position while the absolute value of the bias voltage was increased. We found that decarbonylation was rather delocalized and hence also possible for lateral distances of several nm between tip and molecule. Because of the nonlocality of the process, we tentatively propose that it is mediated by hot interface-state charge carriers (19, 44-46), without excluding an influence of the electric field on the dissociation barriers. A detailed study of the reaction mechanisms of the different induced reactions goes beyond the scope of this paper. For different reactions we observed different voltage thresholds that increased with the number of decarbonylated groups. The voltage threshold for decarbonylation, at positive voltage polarity and typical currents on the order of pA, increased from about +1.5 V for the first decarbonylation ($C_{24}O_6 \rightarrow C_{22}O_4$) to about +2.5 V for the final, third decarbonylation ($C_{20}O_2 \rightarrow C_{18}$).

The most successful conditions for decarbonylation involved large tip-sample distances (small currents, i.e., $I < 1$ pA) and bias voltages of about +3 V. In most cases, two, four or six CO molecules were removed with one voltage pulse, such that either the product cyclo[18]carbon or an intermediate, with an even number of COs still attached, was formed. The most abundantly formed intermediates were those with either one or two intact masking groups still attached.

Other molecules less frequently created from $C_{24}O_6$ by voltage pulses are shown in fig. S13. Less common intermediates in which the AFM image suggests that in two masking groups a single CO unit was removed to generate two monoketenes and subsequent rearrangement of the polyynic region to reveal an additional triple bond are summarized as 'rare intermediates' (fig. S13, A to G). Rare cyclic reaction products that could not be further manipulated are indicated as 'rare final products' (fig. S13, H to M). Reactions in which the cyclic unit was broken are indicated as 'non-cyclic products' (fig. S13, N to O).

**Table S1.**

**Reaction statistics for on-surface reactions by atom manipulation**. Educts were the intact precursor $C_{24}O_6$, the two most abundant intermediates, $C_{22}O_4$ and $C_{20}O_2$ (structures shown in Fig. 2 in the main text) and other seldomly observed intermediates (rare intermediates). Reaction products were the two most abundant intermediates, $C_{22}O_4$ and $C_{20}O_2$, seldomly observed intermediates (rare intermediates), cyclo[18]carbon ($C_{18}$), chain-like structures due to breaking of a bond within the cyclic system (non-cyclic products) and other rare cyclic reaction products that could not be further manipulated (rare final products; see fig. S13). For the table in total $N = 90$ single molecule reactions induced by atom manipulation were evaluated.

| Educt \ Product | $C_{22}O_4$ | $C_{20}O_2$ | rare intermediates | $C_{18}$ | non-cyclic products | rare final products |
|---|---|---|---|---|---|---|
| $C_{24}O_6$ | 17% | 28% | 4% | 4% | 26% | 21% |
| $C_{22}O_4$ | / | 31% | 6% | 13% | 31% | 19% |
| $C_{20}O_2$ | / | / | 19% | 13% | 38% | 31% |
| rare intermediates | / | 20% | / | 20% | 20% | 40% |

Table S1 summarizes the outcome of all tip-induced decarbonylation attempts starting from either the intact precursor or one of the intermediates. The two most abundant intermediates are listed separately and denoted by their chemical formula $C_{24}O_6$ and $C_{20}O_2$, respectively. We evaluated a total of 90 atomic manipulation events, in which the initial and final product were characterized. The resulting yield for the generation of cyclo[18]carbon from the precursor $C_{24}O_6$ by atom manipulation, including its synthesis *via* intermediates, was 13%. This yield was calculated by summing the measured individual yields for the different pathways, shown in

Table S1. The pathways with different intermediates observed and their determined respective yields are:

$C_{24}O_6 \rightarrow C_{18}$ (4% yield),

$C_{24}O_6 \rightarrow C_{20}O_2 \rightarrow C_{18}$ (3.6% yield),

$C_{24}O_6 \rightarrow C_{22}O_4 \rightarrow C_{18}$ (2.2% yield),

$C_{24}O_6 \rightarrow$ rare intermediate $\rightarrow C_{18}$ (0.8% yield),

$C_{24}O_6 \rightarrow C_{22}O_4 \rightarrow C_{20}O_2 \rightarrow C_{18}$ (0.7% yield),

$C_{24}O_6 \rightarrow C_{20}O_2 \rightarrow$ rare intermediate $\rightarrow C_{18}$ (1.1% yield),

$C_{24}O_6 \rightarrow C_{22}O_4 \rightarrow$ rare intermediate $\rightarrow C_{18}$ (0.2% yield),

$C_{24}O_6 \rightarrow$ rare intermediate $\rightarrow C_{20}O_2 \rightarrow C_{18}$ (0.1% yield).

Total yield of $C_{18}$: 12.7%.

## Covalently fused molecules

When voltage pulses are applied to two proximate molecules, molecular fusion can be achieved. We could trigger such reactions by applying bias voltages above or close to two adjacent molecules as described in the main text. The suggested reaction mechanism for the formation of the fused molecule shown in Fig. 4 in the main text is shown in fig. S19. The nonlocality of the process suggests that it is mediated by hot interface-state charge carriers (19, 44-46). Figure S20 depicts the formation by atom manipulation of a large ring with at least one unknown side group. Figure S20A shows an STM overview image before the ring was formed. The employed $C_{24}O_6$ precursor (fig. S20B) is highlighted by a yellow box. Voltage pulses of $V = 2.4$ V and $V = 3.5$ V, respectively, were applied atop the molecule, resulting in partial decarbonylation (fig. S20, C to F). Eventually, the molecule shown in fig. S20, G and H was found in the area marked by the green box in fig. S20A. Upon scanning with an elevated bias voltage ($V = 3.7$ V, $I = 1$ pA), the molecule jumped away. The fused molecule shown in fig. S20, I to N was found close to the molecule marked by the red circle in fig. S20A. Figure S20, I to K show STM images and AFM images at two different tip-sample distances, from which the suggested structure shown in fig. S20L is inferred. HOMO and LUMO densities of the fused cyclic molecule are shown in fig. S20, M and N.

In addition to molecules fused by atom manipulation, we also found fused molecules on the surface which had likely reacted during the sublimation and adsorption process. Figure S21 shows a molecule that was found on the surface, where no voltage pulses had been applied in the vicinity before. The AFM contrast suggests that this molecule was formed by coalescence of three partially decarbonylated precursor molecules. STM orbital density images (fig. S21, C and D) reveal that the π-system is delocalized over the entire molecule.

**Density functional theory calculations**

Density functional theory (DFT) calculations were carried out using the all-electron code FHI-aims (47). Two different approximations to the exchange-correlation (x-c) functional were used; the Perdew–Burke-Ernzerhof (PBE) generalized gradient approximation (48) and the Heyd–Scuseria–Ernzerhof (HSE) hybrid functional (49). The mixing parameter that controls the amount of exact exchange in HSE was set to 0.8. The Tkatchenko-Scheffler correction (50) was used to account for van der Waals interactions. For calculations using PBE, we employed the default 'very tight' settings for the atomic basis sets, while for HSE the default 'tight' settings were employed. All molecules were calculated in the gas phase, and the structures were relaxed until the forces acting on atoms were smaller than $10^{-3}$ eVÅ$^{-1}$.

We obtained different results using the two different x-c functionals. Independent of the starting geometry, the PBE calculations for cyclo[18]carbon converged to a cumulenic structure with 18-fold symmetry and equal bond length of all C=C bonds of 1.284 Å. In contrast, the HSE calculations for cyclo[18]carbon always converged to a 9-fold symmetry with alternating short (1.195 Å) and long (1.343 Å) bonds, i.e., a polyynic structure.

For the intermediates ($C_{22}O_4$, $C_{20}O_2$) we obtained different results with the two x-c functionals as well. The HSE functional yielded similar bond-length alternation (BLA) for the different length polyynic segments in the different molecules ($C_{24}O_6$, $C_{22}O_4$, $C_{20}O_2$, $C_{18}$) and larger BLA compared to calculations with PBE. For the PBE functional, we observed that the BLA monotonically decreases with the length of the polyynic chain in the molecules $C_{24}O_6$, $C_{22}O_4$, $C_{20}O_2$ and vanishes for cyclo[18]carbon.

**AFM image simulations**

AFM images acquired with CO-functionalized tips can be efficiently simulated *via* a well-established molecular-mechanics model that takes into account the flexibility of the CO molecule at the tip apex (25, 31). The interaction of the tip with the sample is modelled *via* classical force fields and the CO at the tip is relaxed. We simulated AFM images for the precursor $C_{24}O_6$, the partially decarbonylated intermediates ($C_{22}O_4$, $C_{20}O_2$) as well as the final product cyclo[18]carbon using the Probe-Particle code implemented by Hapala *et al.* (25). This code also allows electrostatic interactions to be included by assigning a charge to the CO molecule and letting it interact with the Hartree potential of the sample.

As input for the AFM simulation, we used the atomic coordinates from the DFT gas-phase structure optimization as well as the corresponding DFT Hartree potential. For all AFM simulations, we used a lateral spring constant for the CO at the tip of 0.2 N/m and an electrostatic monopol on the oxygen of –0.05 e (51) and the experimental oscillation amplitude $A$ = 50 pm.

Figure S15 shows simulated AFM images for the various molecules at different tip-sample distances, where $\Delta z$ is defined as in the main text, i.e., positive $\Delta z$ values correspond to an increase in the tip-sample distance. All images have been calculated using the results obtained from the HSE x-c functional as input, except the bottom row which was calculated based on the PBE x-c functional results. The (HSE-based) simulated images for the pristine precursor $C_{24}O_6$ (fig. S15A and Fig. 3, D and E in the main text) show very good agreement with the experimental images, with two bright lobes on each of the three sides of the triangular carbon backbone. Each of the bright lobes can be ascribed to one of the triple bonds. For the two intermediates $C_{22}O_4$ and $C_{20}O_2$ (fig. S15B and S15C, respectively) the agreement with experimental data is very good as well. At large and medium tip-sample distances ($\Delta z > 0.3$ Å), the AFM simulations show five bright lobes at the longer bend part of the molecule for $C_{22}O_4$ and eight lobes for $C_{20}O_2$, which again correspond to the triple bonds. Because for the AFM simulations the optimized gas-phase geometry was used as an input, the simulation does not account for non-planar adsorption geometries. Therefore, all sides of the molecule exhibit equal contrast. Additionally, the CO moieties appear brighter compared to the experimental AFM images since they are not bent down towards the surface in the geometry that was used for simulating the AFM images.

The simulated HSE-based AFM images for the final product cyclo[18]carbon (fig. S15D, and Fig. 3, S and T in the main text) show nine lobes at large and medium tip-sample distances (see fig. S15D, $\Delta z > 0.3$ Å), which form a ring and can be associated with the nine triple bonds of the polyynic structure. At small tip-sample distances (see fig. S15D, $\Delta z < 0.3$ Å), the AFM image of the molecule resembles a nonagon due to bending of the CO at the tip apex (25, 31) (a similar change in contrast at small tip-sample distances occurs for the other molecules shown in fig. S15, A to C). The AFM contrast for all molecules and all tip heights, using the results of the HSE x-c functional as inputs, are in excellent agreement with the experimental observations. Again, the simulation does not account for a non-planar adsorption geometry. Hence, the differences in brightness between two opposing sites of the molecules resulting from the non-planarity of the molecule with respect to the surface are not reproduced in the simulations.

For comparison, fig. S15E shows simulated AFM images based on the results obtained with the PBE x-c functional, i.e., for cumulenic cyclo[18]carbon. They show only a featureless ring, independent of the tip-sample distance, in contrast to the experimental images. Thus, the AFM simulations confirm our interpretation of the experimental data that cyclo[18]carbon adopts the polyynic structure on NaCl.

## Selected NMR Spectra

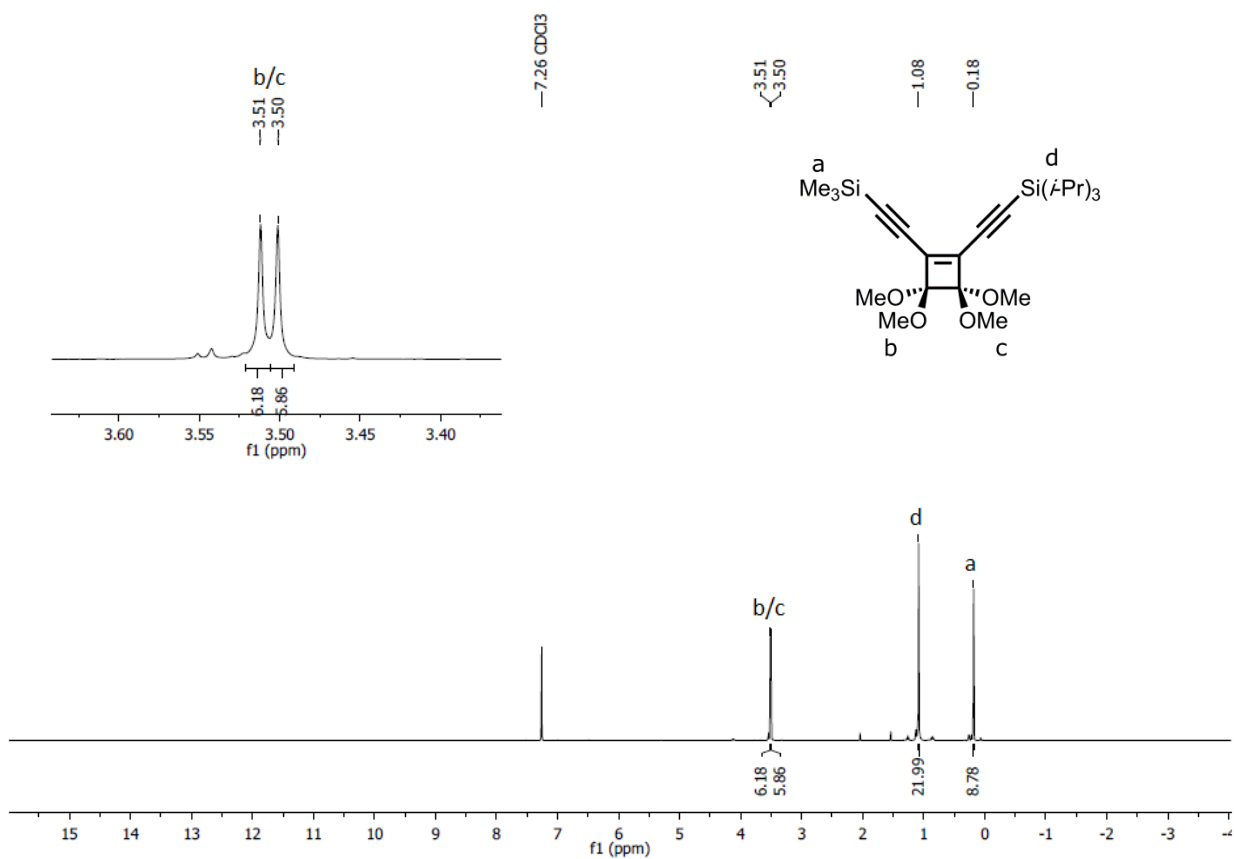

**Fig. S1.**
$^1$H NMR spectrum of **S3** (400 MHz, CDCl$_3$, 21 °C).

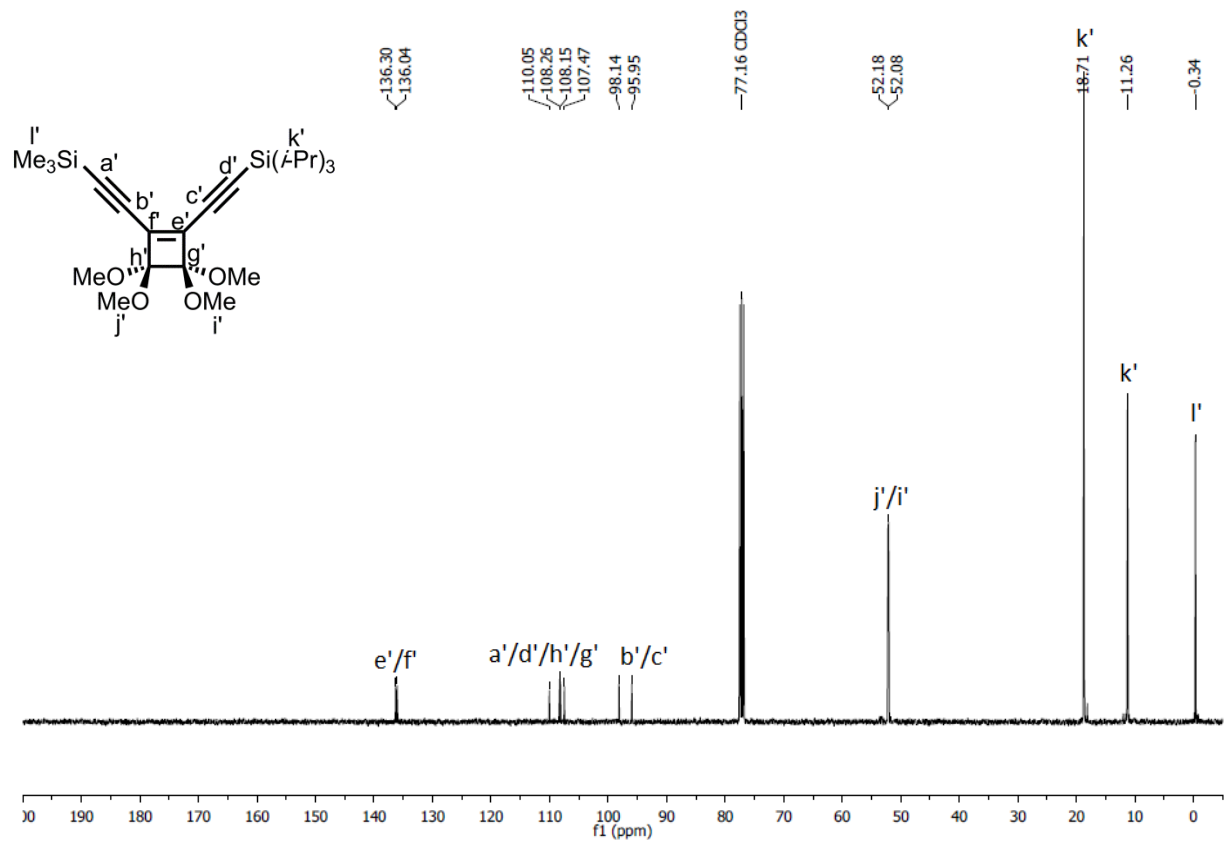

**Fig. S2.**
$^{13}$C NMR spectrum of **S3** (101 MHz, CDCl$_3$, 21 °C).

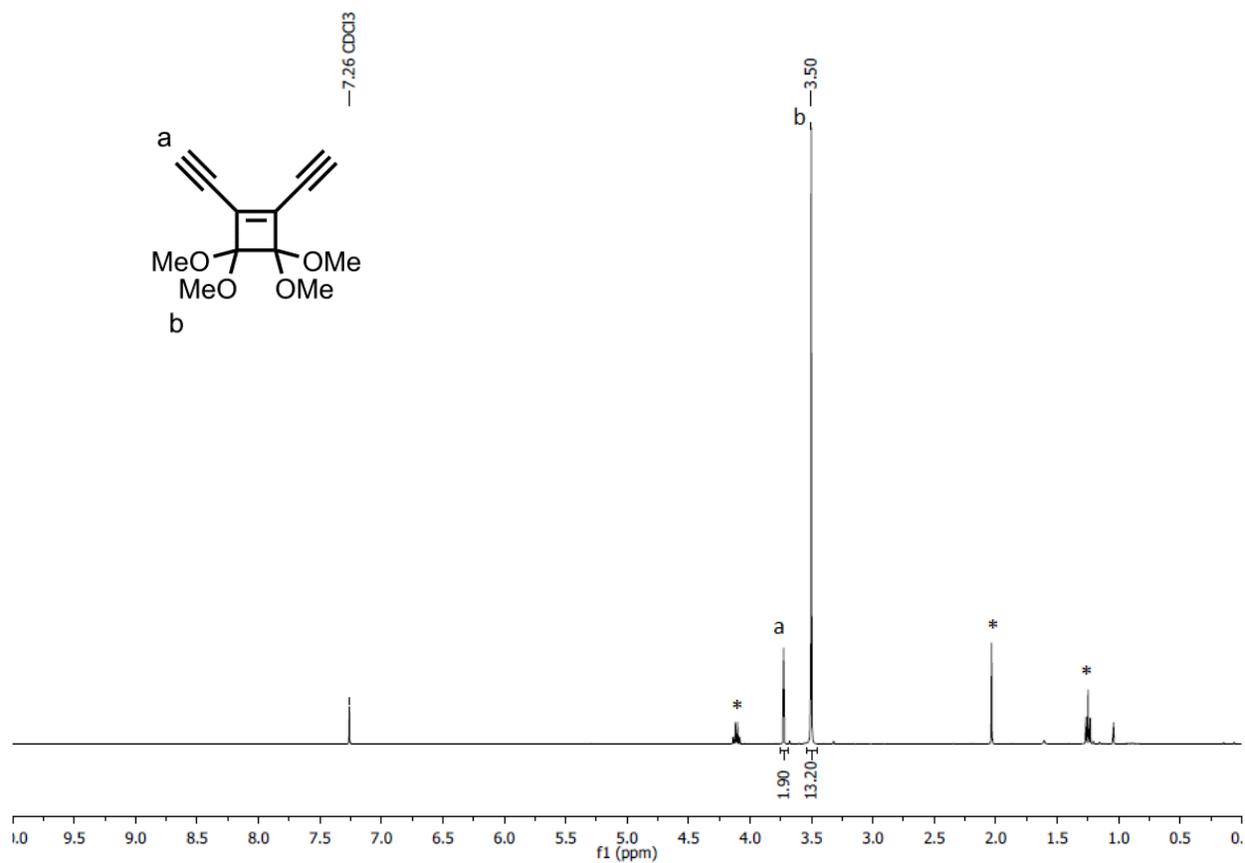

**Fig. S3.**
¹H NMR spectrum of **S4** (400 MHz, CDCl$_3$, 21 °C). Residual ethyl acetate solvent signals (*) as samples were not completely dried to avoid decomposition. Reaction yield was corrected accordingly.

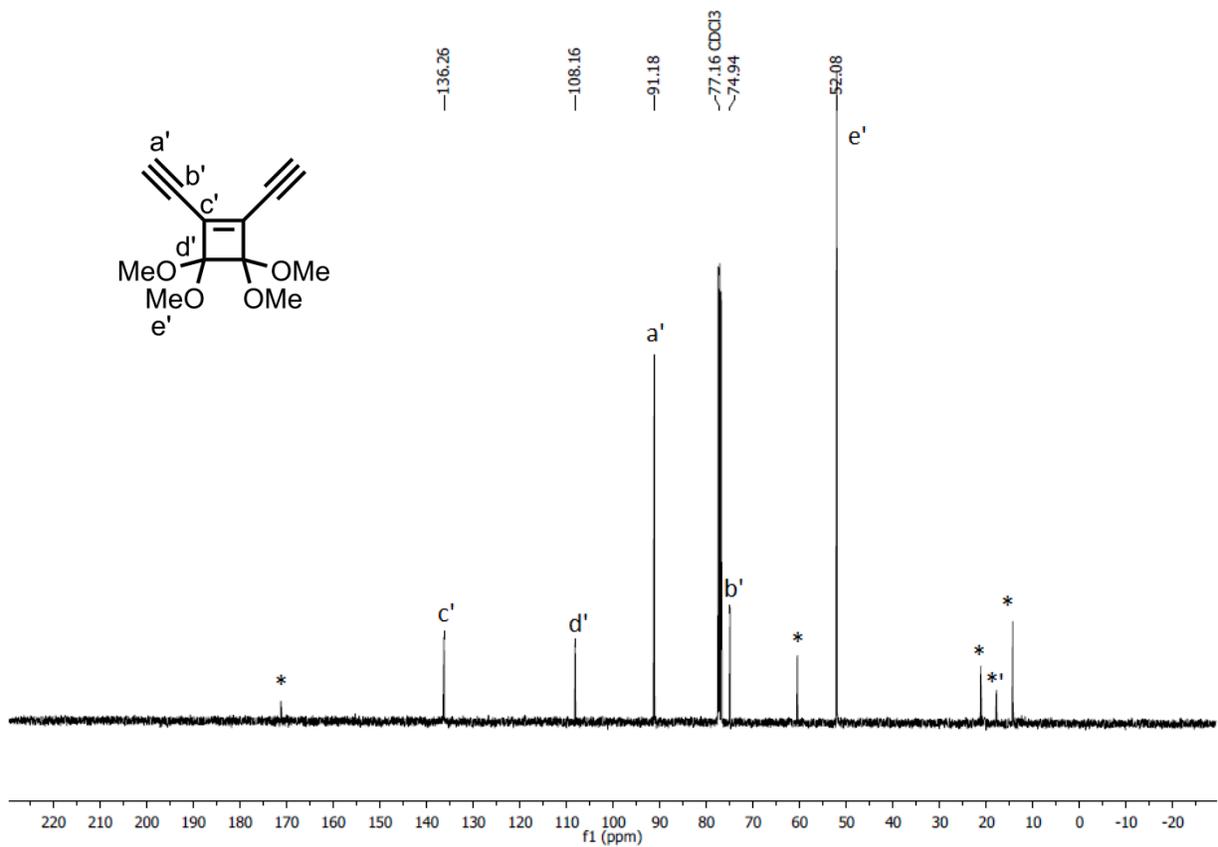

**Fig. S4.**
[13]C NMR spectrum of **S4** (101 MHz, CDCl$_3$, 21 °C). Residual ethyl acetate solvent signals (*) as samples were not completely dried to avoid decomposition. Reaction yield was corrected accordingly.

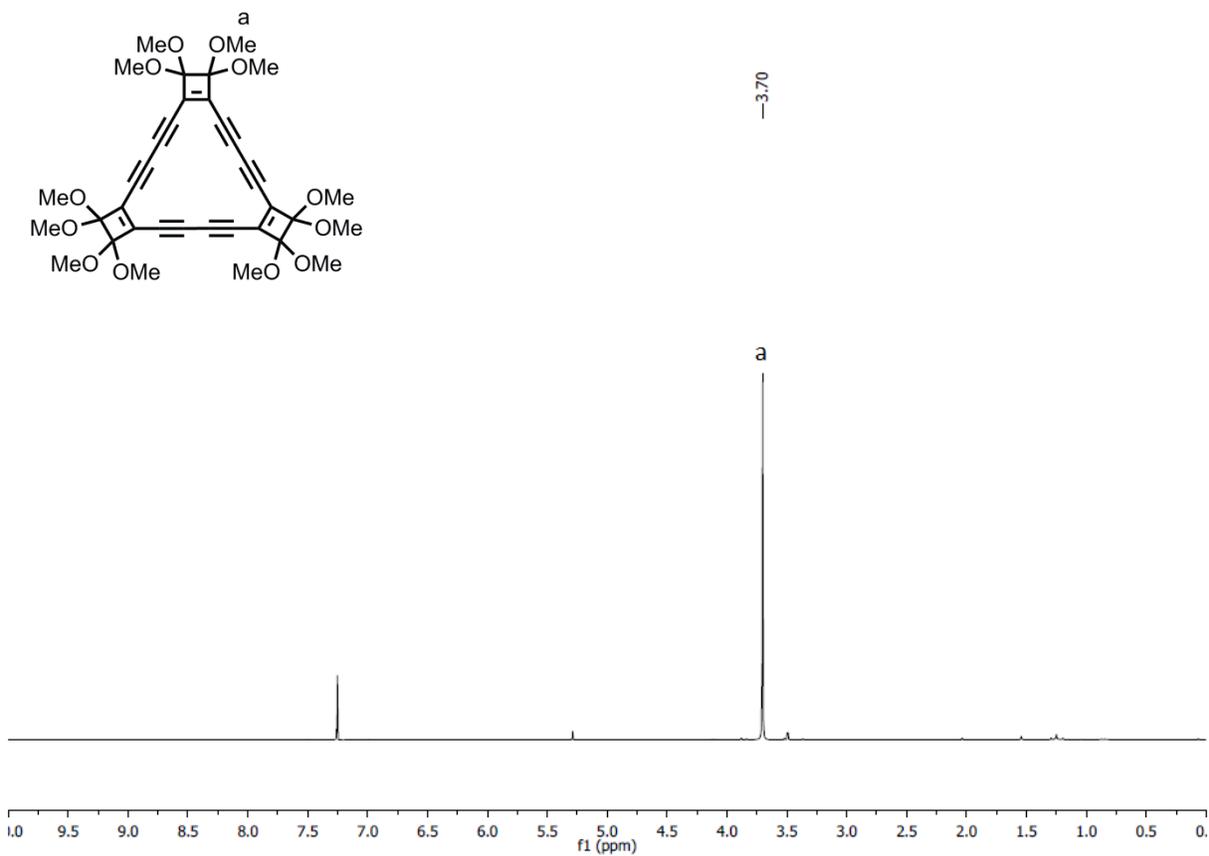

**Fig. S5.**
¹H NMR spectrum of **S5** (400 MHz, CDCl₃, 21 °C).

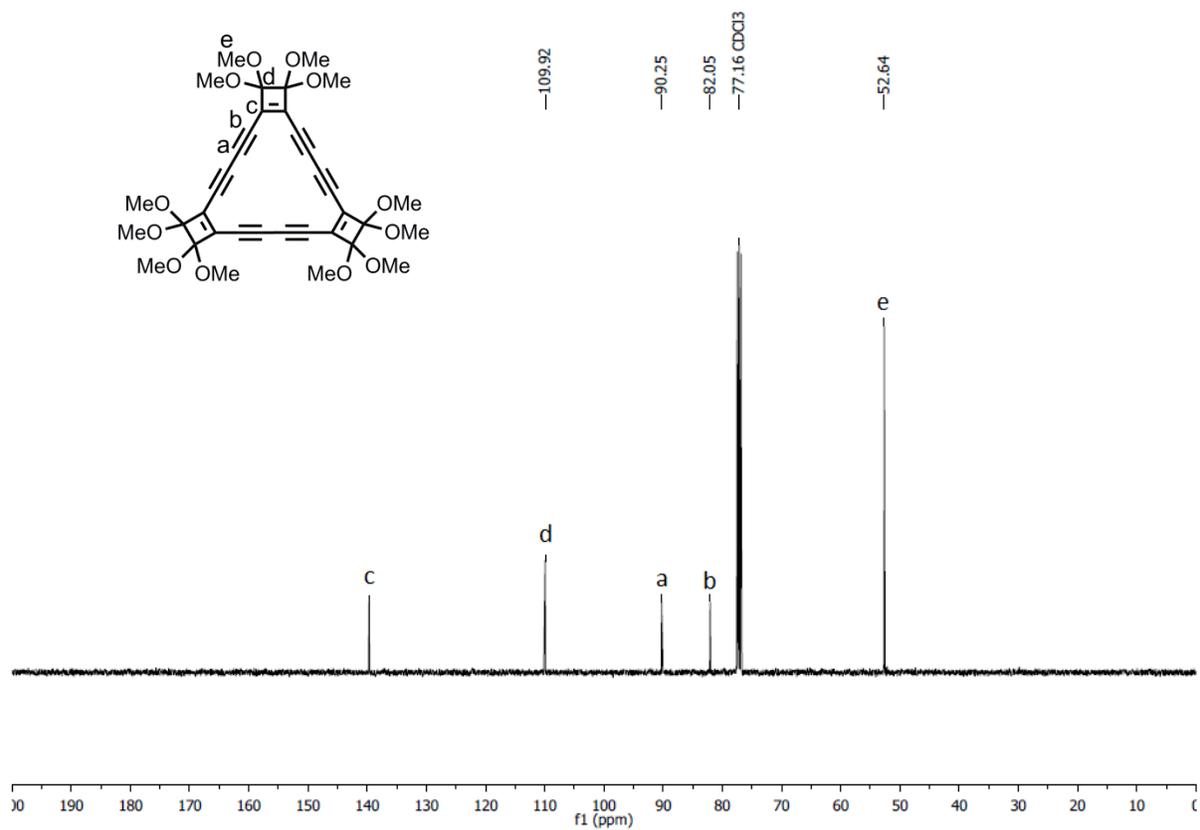

**Fig. S6.**
$^{13}$C NMR spectrum of **S5** (101 MHz, CDCl$_3$, 21 °C).

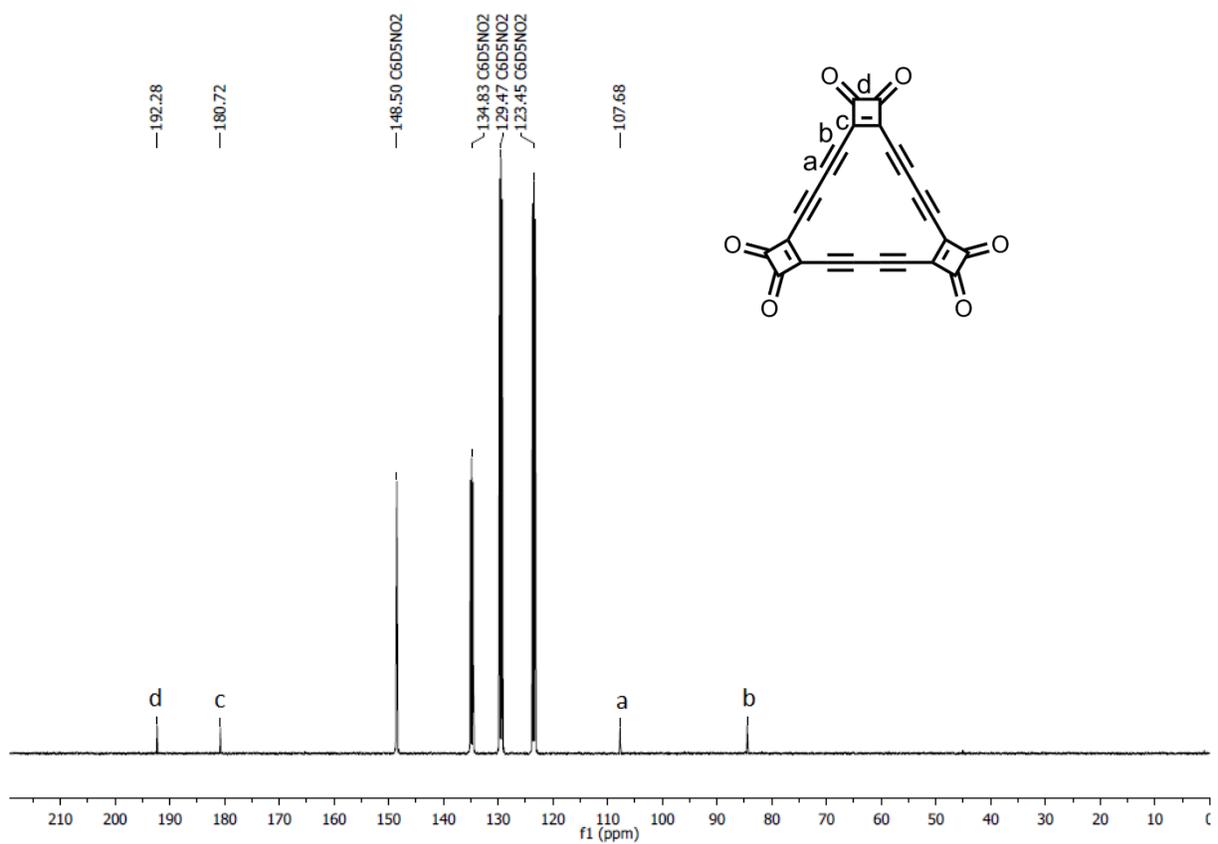

**Fig. S7.**
$^{13}$C NMR spectrum of **S6**, $C_{24}O_6$ (101 MHz, $C_6D_5NO_2$, 21 °C)

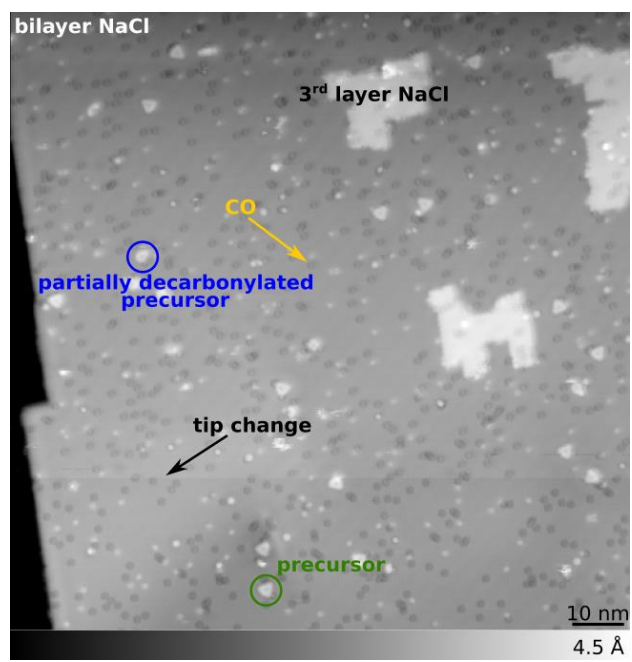

**Fig. S8.**

**Overview image.** STM overview recorded at a set point of $V = 0.4$ V, $I = 0.5$ pA. Most of the image shows bilayer NaCl on Cu(111), a few islands of third layer NaCl are visible (3$^{rd}$ layer NaCl). The small depressions are single CO molecules. The overview shows mainly intact $C_{24}O_6$ precursor molecules (example shown in the green circle) and a few partially decarbonylated intermediates (example shown in the blue circle). In the lower part of the image, the contrast changes abruptly, which can be attributed to a tip change.

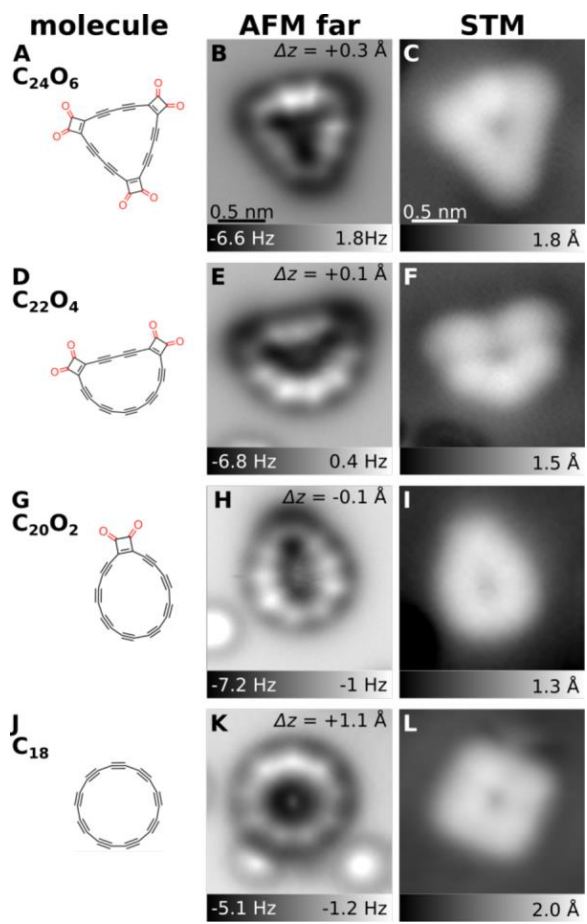

**Fig. S9.**

**STM contrast of precursors and products.** The structures (1st column) and AFM data (2nd column) are reproduced from Fig. 3 of the main text. The 3rd column shows complementary STM data of the same molecules in that row, respectively. STM images were recorded in constant-current mode at $I = 0.5$ pA, $V = 0.2$ V (C, F, I) and $I = 0.7$ pA, $V = 0.1$ V (L). The same scale bar as in (B) and (C) applies to all AFM and all STM images, respectively. All measurements are obtained with CO-functionalized tips.

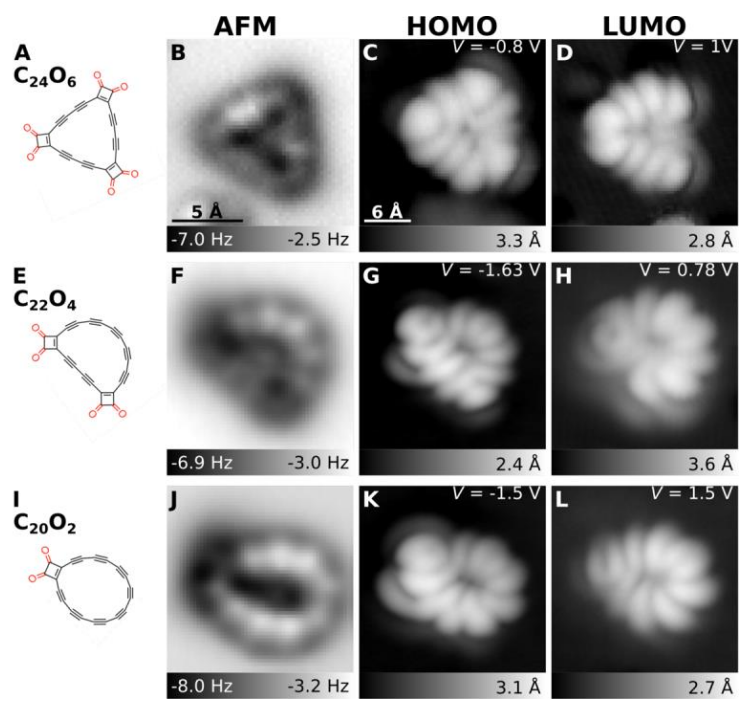

**Fig. S10.**

**Orbital density maps**. Molecular structures (first column), AFM images (second column) and STM orbital density images (third and fourth column) of the cyclo[18]carbon precursors (**A–D**) $C_{24}O_6$ (STM at $I = 1$ pA), (**E–H**) $C_{22}O_4$ (STM at $I = 0.5$ pA) and (**I–L**) $C_{20}O_2$ (STM at $I = 0.5$ pA). The images acquired at the first resonance at negative bias voltage are interpreted as density of the highest occupied molecular orbital (HOMO), while the images acquired at the first resonance at positive bias voltage are interpreted as density of the lowest unoccupied molecular orbital (LUMO).

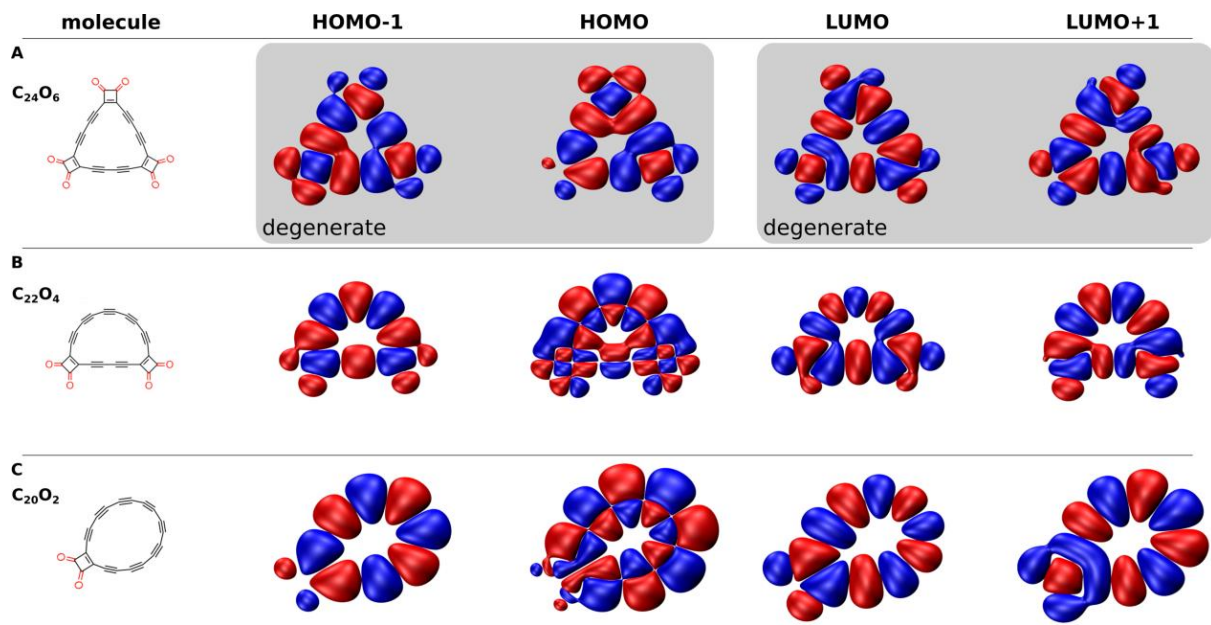

**Fig. S11.**

**Calculated orbital density maps**. Electron densities of the two highest occupied molecular orbitals (HOMO-1 and HOMO) and lowest unoccupied molecular orbitals (LUMO and LUMO+1) of the cyclo[18]carbon precursors (**A**) $C_{24}O_6$, (**B**) $C_{22}O_4$ and (**C**) $C_{20}O_2$ in the gas phase. The DFT calculations were carried out using the HSE hybrid functional.

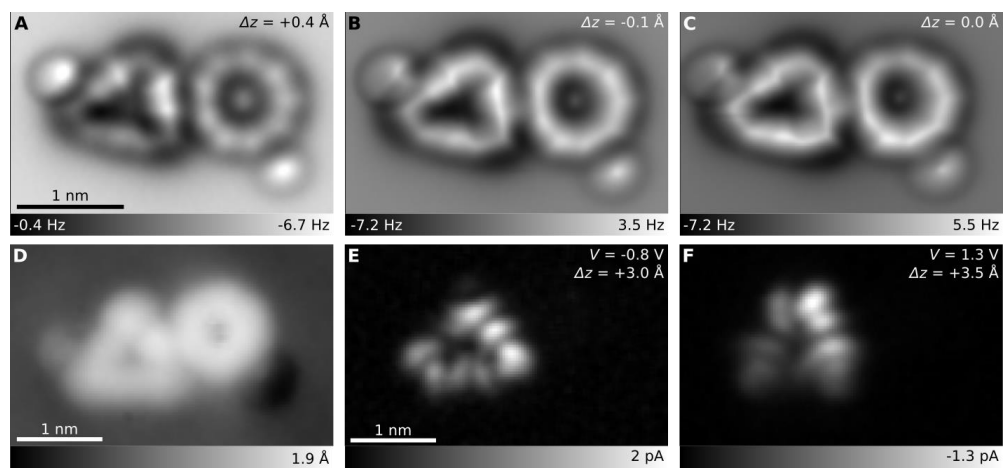

**Fig. S12.**

**Comparative measurements of $C_{24}O_6$ precursor and cyclo[18]carbon**. (**A–C**) AFM and (**D–F**) STM images of a $C_{24}O_6$ precursor and a cyclo[18]carbon molecule adsorbed adjacent to each other. The bright features in the AFM images next to the molecule stem from adjacently adsorbed CO molecules. (D) In-gap constant-current STM image at $V = 0.13$ V and $I = 0.5$ pA. (E, F) Constant-height STM images obtained at the onset of the PIR and NIR of the $C_{24}O_6$ precursor, respectively. In the images recorded in constant-height mode (A–C, E, F), $\Delta z$ corresponds to the tip-height offset with respect to an STM set point of $V = 0.2$ V and $I = 0.5$ pA.

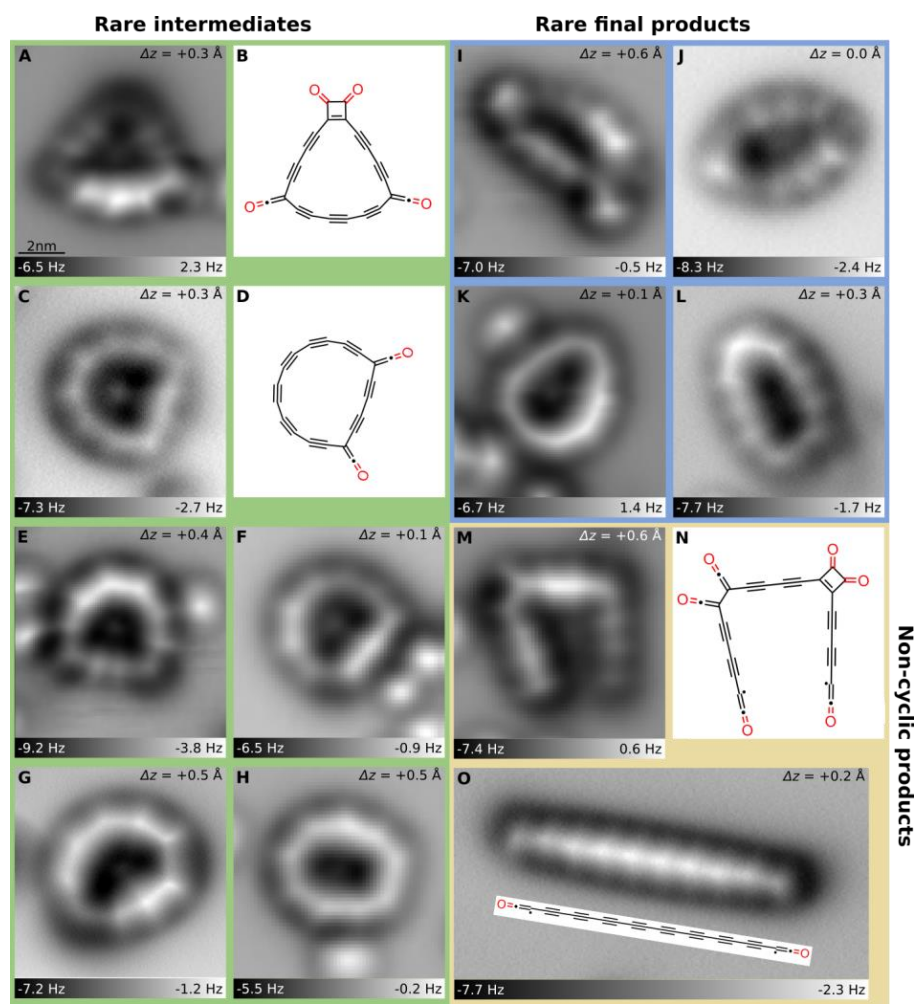

**Fig. S13.**
**Molecules less commonly formed by atom manipulation**. (**A, C, E–M, O**) AFM images and (**B, D, N, O**) chemical structures. (**A–H**) Less common intermediates, summarized as 'rare intermediates' in the reaction statistics (Table S1). B and D show the assigned structure of the molecules shown in A and C. In these the removal of 2 CO and 4 CO from $C_{24}O_6$, generated different $C_{22}O_4$ and $C_{20}O_2$ isomers compared to the ones described in the main text. (**I–L**) AFM images of cyclic molecules that did not react further upon voltage pulses. In the reaction statistics, molecules of this kind are summarized as 'rare final products'. (**M–O**) Products in which a bond within the cyclic unit was broken. (N) shows the assigned structure to the molecule shown in M. (O) shows an AFM image of a linear polyyne chain with the assigned structure shown in the inset. In the reaction statistics, molecules of this kind are summarized as 'non-cyclic products'. $\Delta z$ is given with respect to the STM set point of $V = 0.2$ V and $I = 0.5$ pA in A, F, H, K and O; $V = 0.2$ V and $I = 0.8$ pA in C and L; $V = 0.15$ V and $I = 0.6$ pA in E; $V = 0.2$ V and $I = 1$ pA in G; $V = 0.1$ V and $I = 0.7$ pA in I; $V = 0.3$ V and $I = 0.5$ pA in J; $V = 0.1$ V and $I = 0.5$ pA in M.

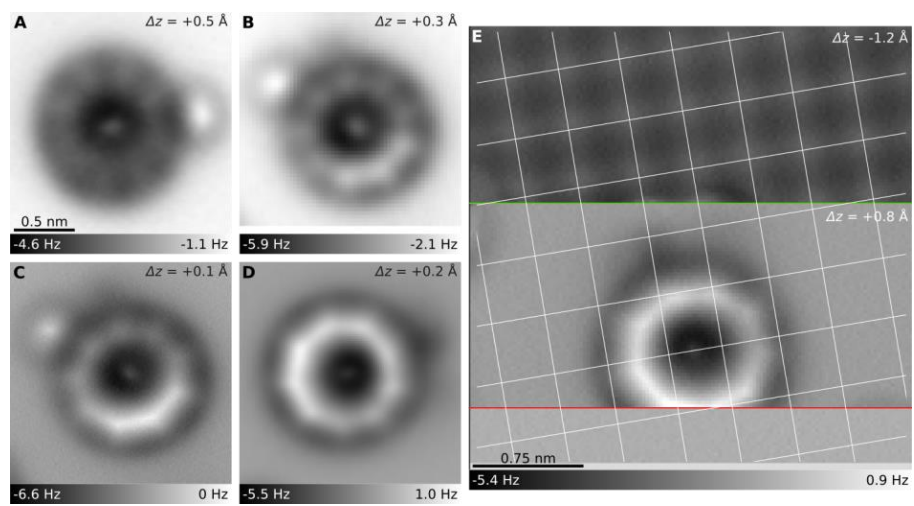

**Fig. S14.**
**Additional AFM data of cyclo[18]carbon.** (**A–D**) Constant-height AFM images of different cyclo[18]carbon molecules recorded at different tip-sample distances, decreasing from (A) to (D). (B) and (C) show the same molecule. (**E**) AFM image of cyclo[18]carbon to determine the adsorption position. The slow scan direction was from top to bottom. In the upper part of the image, a small tip height ($\Delta z = -1.2$ Å) was used to resolve the NaCl surface atomically. The Cl anions appear as bright features and are indicated by the overlaid grid. At the green line, the tip height was increased to $\Delta z = 0.8$ Å. At the red line, the tip lost the CO molecule. The measurement shows cyclo[18]carbon adsorbed on a Cl-Cl bridge site. In (A, D, E), $\Delta z$ is given with respect to an STM set point of $V = 0.1$ V and $I = 0.5$ pA, in (B, C), $\Delta z$ is given with respect to an STM set point of $V = 0.2$ V and $I = 0.5$ pA.

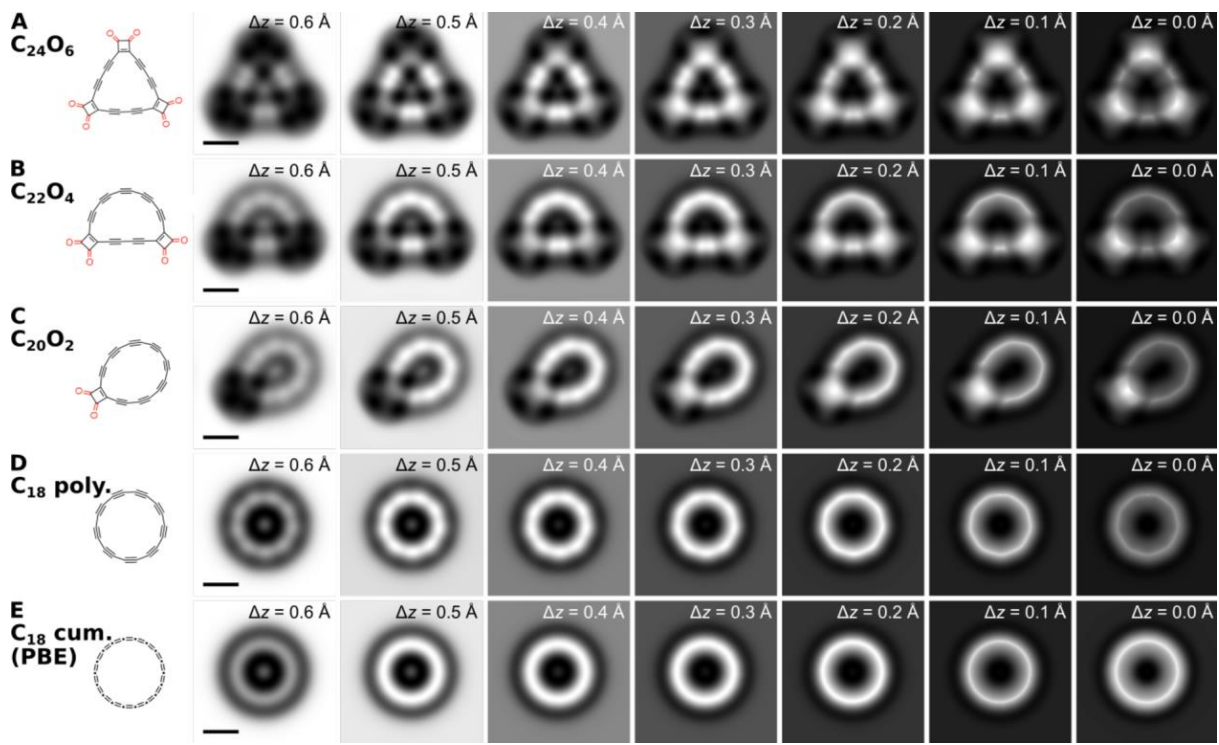

**Fig. S15.**

**AFM simulations**. Chemical structure and simulated AFM images at different tip-sample distances for the pristine precursor $C_{24}O_6$ (**A**), the intermediates $C_{22}O_4$ (**B**) and $C_{20}O_2$ (**C**) as well as polyynic cyclo[18]carbon (**D**) and cumulenic cyclo[18]carbon (**E**). The molecular structure was optimized by DFT using either the HSE x-c functional (A–D) or the PBE x-c functional (E). $\Delta z$ values correspond to the increase in tip-sample distance. The scale bar is 5 Å and all images have the same scale.

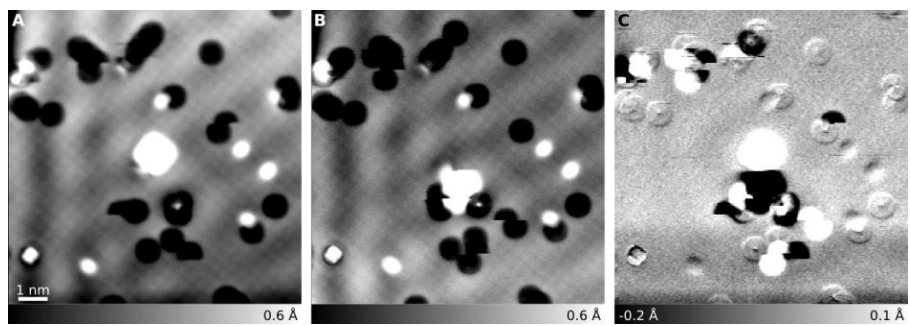

**Fig. S16.**
**Interface state scattering behavior of neutral cyclo[18]carbon.** (**A, B**) STM images acquired at $I = 0.7$ pA, $V = 0.1$ mV before and after moving the (charge neutral) cyclo[18]carbon, in the center of the image, laterally on the surface, respectively. (**C**) Difference image (A) – (B), exhibiting almost no contrast of standing wave pattern.

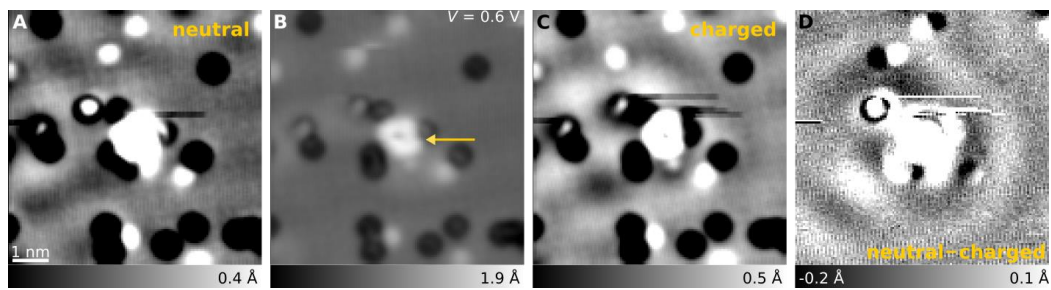

**Fig. S17.**

**Interface state scattering behavior of charged cyclo[18]carbon.** (**A**) Image of the neutral cyclo[18]carbon acquired at $I$ = 0.4 pA, $V$ = 0.1 V. (**B**) Image of the same cyclo[18]carbon acquired at $I$ = 0.4 pA, $V$ = 0.6 V, leading to a change in charge state at the scan line indicated by the arrow. (**C**) Image of the charged cyclo[18]carbon acquired at $I$ = 0.4 pA, $V$ = 0.1 V. (**D**) Difference image of (A) – (C) showing an interference pattern, indicating charge-state switching of the molecule.

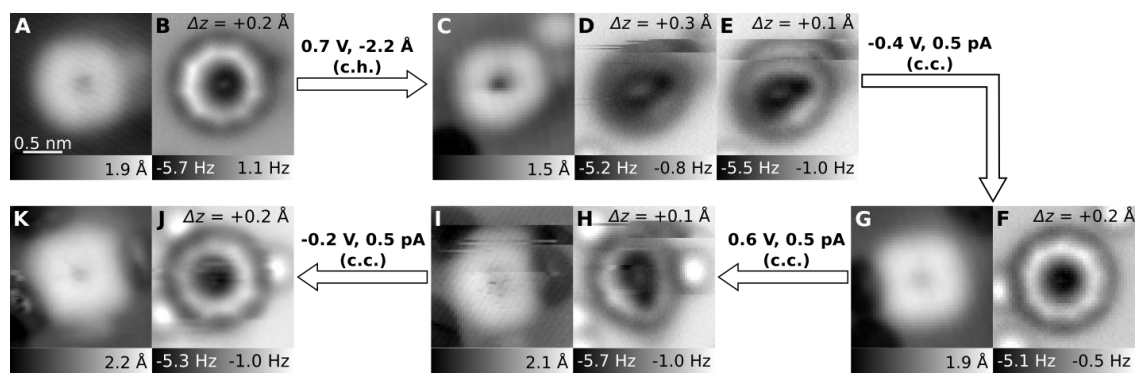

**Fig. S18.**
**Reversible charge-state transitions of cyclo[18]carbon.** Series of STM (**A, C, F, I, K**) and AFM (**B, D, E, G, H, J**) images recorded after charging and discharging events of a cyclo[18]carbon molecule, starting with the neutral molecule (A, B). Upon applying positive bias voltages $V > 0.6$ V, the molecule could be charged (C–E), upon applying negative voltages the molecule could be switched back into its neutral state (G, F). It was charged again (I, H) and switched back to neutral again (K, J). The arrows indicate how the charge state was manipulated, 'c.h.' indicates that the change in charge state happened during a constant-height image with the given parameters, 'c.c.' indicates that the change occurred during a constant-current STM image at the given parameters. Note the markedly different, less symmetric and less stable AFM contrast of the charged cyclo[18]carbon compared to the neutral one. STM images displayed were recorded at a set point of $V = 0.1$ V and $I = 0.5$ pA and $\Delta z$ for the constant-height AFM images displayed is given with respect to that set point.

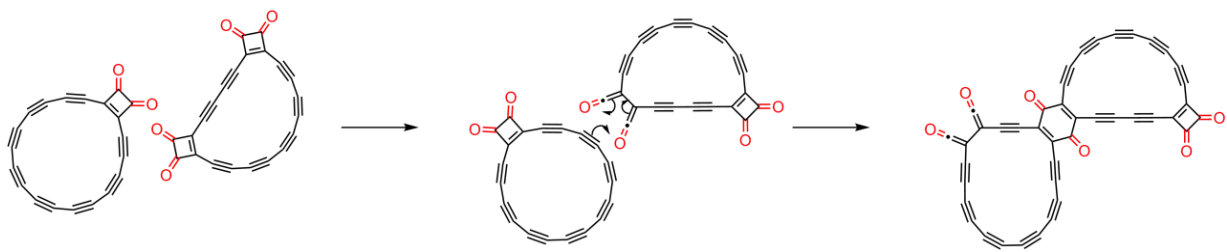

**Fig. S19.**

**Possible mechanism for the reaction shown in Fig. 4.** Possible mechanism for the fusion reaction between $C_{20}O_2$ and $C_{22}O_4$, shown in Fig. 4 of the main text. One of the cyclobutene-1,2-dione groups of $C_{22}O_4$ undergoes a [2+2] cycloreversion to form a bisketene moiety. This then undergoes a [4+2] cycloaddition with a triple bond of $C_{20}O_2$ to form the benzoquinone product.

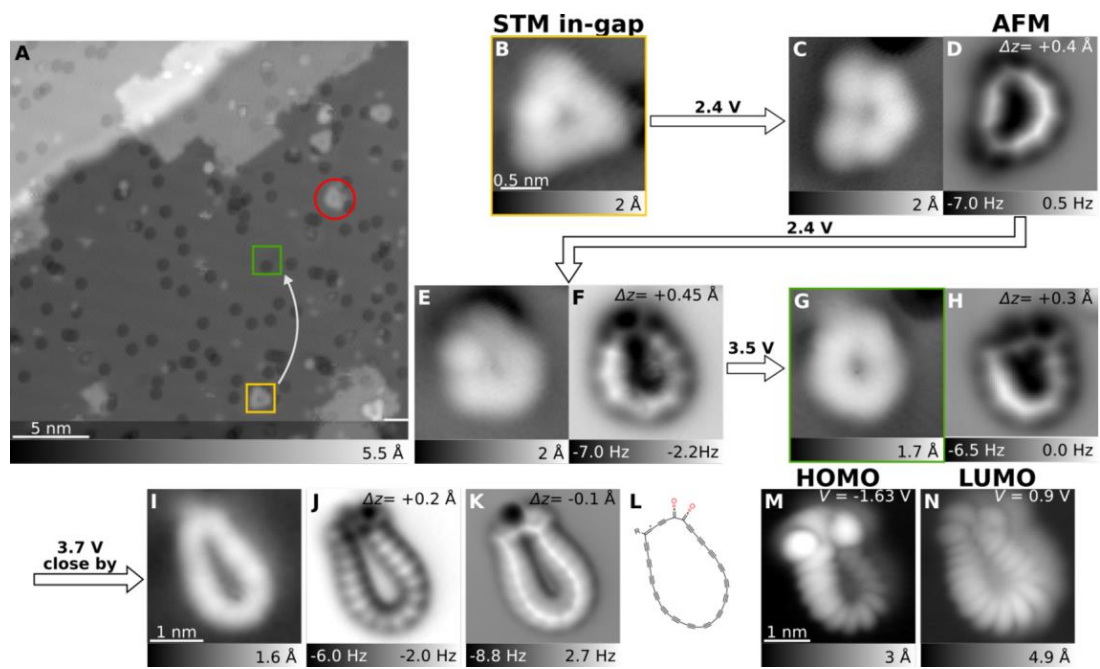

**Fig. S20.**
**Pathway for the reaction of two adjacent molecules.** (**A**) STM overview ($V = 0.6$ V and $I = 0.5$ pA). The yellow box marks the $C_{24}O_6$ precursor, shown in (**B**) that was chemically altered by applying voltage pulses/scanning at elevated bias voltages. Upon applying elevated bias voltages, the molecule was partially decarbonylated, as indicated by the consecutive STM and AFM images (**C–H**). The voltages used to chemically alter the molecule are indicated above the arrows. During decarbonylation, the molecule jumped several times and eventually landed in the area marked by the green box in (A). In the last step, the molecule jumped and most likely reacted with the molecule marked by the red circle in (A), yielding the molecule shown in (**I–K**). (**L**) Suggested structure and (**M**, **N**) STM orbital density maps recorded at the PIR and NIR (at $I = 1$ pA), respectively, of the molecule shown in (I–K). The scale bars in (B, I, M) apply to all following images, respectively. (B–F) and (I–K) were recorded using a set point of $V = 0.2$ V and $I = 0.5$ pA, for (G–H) $V = 0.2$ V, $I = 1$ pA was used.

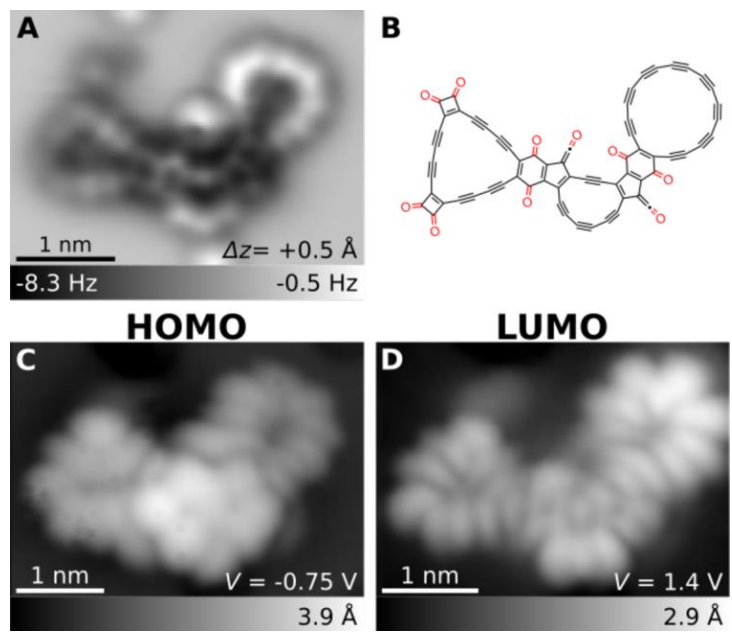

**Fig. S21.**

**Larger fused molecule.** (**A**) AFM image of molecule found like this on the surface. $\Delta z$ is given with respect to the STM set point of $V = 0.2$ V and $I = 0.7$ pA. (**B**) Tentatively assigned structure of the molecule with a molecular formula of $C_{64}O_{10}$. The molecule could have resulted from the reaction $C_{24}O_6 + C_{20}O_2 + C_{20}O_2$. (**C, D**) STM orbital density images of the same molecule recorded at the PIR and NIR, respectively, at $I = 0.7$ pA. Delocalization of the orbital densities over the entire molecule indicate a delocalized π-system.